\documentclass[a4paper, 11pt]{article}
\usepackage[utf8]{inputenc}
\usepackage{cite}
\usepackage{graphicx}
\usepackage{xcolor}

\title{ New combinational therapies for cancer using modern statistical mechanics}
\author{Jorge A. González$^1$\footnote{jorgalbert3047@hotmail.com}, M. Acanda$^2$, Z. Akhtar$^3$, D. Andrews$^4$, J. I. Azqueta$^2$, E. Bass$^4$,\\ A. Bellorín$^5$, J. Couso$^8$,
Mónica A. García-Ñustes$^9$, Y. Infante$^8$,\\ S. Jim\'enez$^{10}$, L. Lester$^6$, L. Maldonado$^4$, Juan F. Marín$^9$\footnote{juan.marin.m@mail.pucv.cl}, L. Pineda$^7$, \\
I. Rodríguez$^7$, C. C. Tamayo$^2$, D. Valdes$^2$, L. Vázquez$^{11}$.\\\\
{\small $^1$ Department of Physics, Florida International University, Miami, Florida 33199, U.S.A.}\\
{\small $^2$ Department of Biological Sciences, Florida International University, Miami, Florida 33199, U.S.A.}\\
{\small $^3$ Department of Biology, College of Arts and Sciences, University of Miami, Coral Gables, Florida 33146, U.S.A.}\\
{\small $^4$ Medical Campus, Miami Dade College, 950 NW 20$^{th}$ Street, Miami, Florida 33127, U.S.A.}\\
{\small $^5$ Escuela de F\'isica, Facultad de Ciencias, Universidad Central de Venezuela,}\\{\small Apartado Postal 47586, Caracas 1041-A, Venezuela}\\
{\small $^6$ Sylvester Comprehensive Cancer Center, University of Miami Health System,}\\ {\small 1475 NW 12$^{th}$ Ave.,1$^{st}$ Floor,
Miami, Florida 33136, U.S.A.}\\
{\small $^7$ Millner School of Medicine, University of Miami, 1600 NW 10$^{th}$ Ave., 1140, Miami, Florida 33136, U.S.A.}\\
{\small $^8$ College of Engineering and Computing, The Engineering Center, Florida International University,}\\
{\small 10555 West Flagler Street, Miami, Florida 33174, U.S.A.}\\
{\small $^9$ Instituto de F\'isica, Pontificia Universidad Cat\'olica de Valpara\'iso, Casilla 4059, Chile}\\
{\small $^{10}$ Departamento de Matem\'atica Aplicada a las TT.II., E.T.S.I. Telecomunicaci\'on,}\\ {\small Universidad Polit\'ecnica de Madrid, 28040-Madrid, Spain}\\
{\small $^{11}$ Departamento de Matem\'atica Aplicada, Facultad de Inform\'atica,}\\{\small Universidad Complutense de Madrid, 28040-Madrid, Spain}}

\begin{document}

\maketitle

\begin{abstract}

We investigate a new dynamical system that describes tumor-host interaction. The equation that describes the untreated tumor growth is based on
non-extensive statistical mechanics. Recently, this model has been shown to fit successfully exponential, Gompertz, logistic, and power-law
tumor growths. We have been able to include as many hallmarks of cancer as possible. We study also the dynamic response of cancer under therapy.
Using our model, we can make predictions about the different outcomes when we change the parameters, and/or the initial conditions. We can
determine the importance of different factors to influence tumor growth. We discover synergistic therapeutic effects of different treatments
and drugs. Cancer is generally untreatable using conventional monotherapy. We consider conventional therapies, oncogene-targeted therapies,
tumor-suppressors gene-targeted therapies, immunotherapies, anti-angiogenesis therapies, virotherapy, among others. We need therapies with the
potential to target both tumor cells and the tumors' microenvironment. Drugs that target oncogenes and tumor-suppressor genes can be effective
in the treatment of some cancers. However, most tumors do reoccur. We have found that the success of the new therapeutic agents
can be seen when used in combination with other cancer-cell-killing therapies. Our results have allowed us to design a combinational therapy
that can lead to the complete eradication of cancer.

\end{abstract}

\section{Introduction}
\label{Sec:Introduction}

There are many papers dedicated to the mathematical modeling of tumor growth \cite{Norton1976, Lamerton1966, Steel1989, Steel2002,
Sullivan1972, Wheldon1988, Laird1964, Skipper1971, McCredie1965, Norton1977, Norton1986, Hart1998, Calderon1991, Stepanova1979, deVladar2004,
gonzalez2003, gonzalez2006, Lefever1979, Kuznetsov1994, Kuznetsov1996, dOnofrio2005, Eftimie2011, Dingli2006, Albano2006, dOnofrio2010,
Tham2008, Donofrio2006, Donofrio2008, Lo2007, Donofrio2007, Ledzewicz2012, Demidenko2006, Caravagna2010, Albano2011, Cattani2008,
Ledzewicz2013, Wilkie2013, Lithgow2006c, Menchon2007, Kozusko2007, dOnofrio2012, ochab2006coexistence, Ledzewicz2011,
Jimenez2011, Albano2012, dOnofrio2013, Schattler2016, Iwami2012, Roesch2014,Ledzewicz2013-2, Caravagna2012, Sahoo2010, Youshan2007,
Frascoli2014, Moummou2014, Ren2012, Isaeva2009, Patanarapeelert2011, Ledzewicz2014, Albano2007, Ledzewicz2014-2, Ledzewicz2015, Spina2014,
Ledzewicz2014-3, dePillis2014, RomanRoman2012, Albano2008, Chis2009, Ledzewicz2010, Brutovsky2008, Ramos2013, Arshad2016, Albano2015,
Ledzewicz2012-2, Gutierrez2012, Tabatabai2013, Bozkurt2014, Ochab2005, Cattani2013, Cattani2010, Anderson2015, Malinzi2014, Cattani2010-2,
Schattler2012, Vadasz2009, Rozova2016, Schattler2015, Bozkurt2015, Tao2010, Chakraborty2018, Dong2014, Spina2015, Sharifi2018,
Cattani2015, Dunia2011-1, Rosch2016, Oana2013, Amini2015, Dunia2011-2, chakraborty2010stochastic, Ivanov2016, Dunia2011-3,
hu2018dynamics, Ochab2005-2, Dunia2011-4, Yousefi2016, Schattler2015cancer,Rezaei2013optimal,wei2017periodically,lo2007functional,
chis2009analysis,Isaeva2010,Ferrer2016modelizacion,Kiakou2014mathematical,Moral2016simulacion,Sbeity2015review,Albano2012therapy,
sahoo2011stochastic,Lo2009,Shakeri2018adaptive,barrea2017optimal,tsallis2009introduction,cabella2011data,barberis2012describing,galvao2008computational,
barberis2015joint,menchon2008regularization,daSilva2013,khordad2016modeling,grau2007non,sotolongo2003behavior}. In this article, we will
investigate a new dynamical system that describes cancer evolution. With the help of our results, we will design new cancer therapies.

Cancer is caused by genetic changes that activate oncogenes or inactivate tumor suppressor genes \cite{felsher2003cancer,jain2002sustained,
beer2004developmental,shachaf2005tumor,shachaf2005rehabilitation}. Cancer is generally untreatable \cite{Norton1976, Lamerton1966, Steel1989,
Steel2002, Sullivan1972, Wheldon1988, Laird1964, Skipper1971, McCredie1965, Norton1977, Norton1986, Hart1998, Calderon1991, Stepanova1979,
deVladar2004, gonzalez2003, gonzalez2006, Lefever1979, Kuznetsov1994, Kuznetsov1996, dOnofrio2005, Eftimie2011, Dingli2006, Albano2006,
dOnofrio2010, Tham2008, Donofrio2006, Donofrio2008, Lo2007, Donofrio2007, Ledzewicz2012, Demidenko2006, Caravagna2010, Albano2011, Cattani2008,
Ledzewicz2013, Wilkie2013, Lithgow2006c, Menchon2007, Kozusko2007, dOnofrio2012, ochab2006coexistence, Ledzewicz2011,
Jimenez2011, Albano2012, dOnofrio2013, Schattler2016, Iwami2012, Roesch2014,Ledzewicz2013-2, Caravagna2012, Sahoo2010, Youshan2007,
Frascoli2014, Moummou2014, Ren2012, Isaeva2009, Patanarapeelert2011, Ledzewicz2014, Albano2007, Ledzewicz2014-2, Ledzewicz2015, Spina2014,
Ledzewicz2014-3, dePillis2014, RomanRoman2012, Albano2008, Chis2009, Ledzewicz2010, Brutovsky2008, Ramos2013, Arshad2016, Albano2015,
Ledzewicz2012-2, Gutierrez2012, Tabatabai2013, Bozkurt2014, Ochab2005, Cattani2013, Cattani2010, Anderson2015, Malinzi2014, Cattani2010-2,
Schattler2012, Vadasz2009, Rozova2016, Schattler2015, Bozkurt2015, Tao2010, Chakraborty2018, Dong2014, Spina2015, Sharifi2018,
Cattani2015, Dunia2011-1, Rosch2016, Oana2013, Amini2015, Dunia2011-2, chakraborty2010stochastic, Ivanov2016, Dunia2011-3,
hu2018dynamics, Ochab2005-2, Dunia2011-4, Yousefi2016, Schattler2015cancer,Rezaei2013optimal,wei2017periodically,lo2007functional,
chis2009analysis,Isaeva2010,Ferrer2016modelizacion,Kiakou2014mathematical,Moral2016simulacion,Sbeity2015review,Albano2012therapy,
sahoo2011stochastic,Lo2009,Shakeri2018adaptive,barrea2017optimal,tsallis2009introduction,cabella2011data,barberis2012describing,galvao2008computational,
barberis2015joint,menchon2008regularization,daSilva2013,khordad2016modeling,grau2007non,sotolongo2003behavior, felsher2003cancer,
jain2002sustained, beer2004developmental,shachaf2005tumor,shachaf2005rehabilitation,tsallis1988possible,
curado1991generalized, tsallis1995statistical,tsallis1998role,botet2001phase,friberg1997growth,luo2009principles,siegel2014cancer,
coit2013melanoma,chapman2011improved,hauschild2012dabrafenib,hodi2010improved,robert2014anti,berger2012melanoma,davies2002mutations,
nikolaev2012exome,ribas2011braf,trudel2014clinical,trunzer2013pharmacodynamic,flaherty2012combined,prickett2009analysis,guo2011phase,kluger2011phase,
schreiber2011cancer,dong2002tumor,wolchok2013nivolumab,beer2011randomized,rosenberg2011durable,draube2011dendritic,leonhartsberger2012quality,
huber2012interdisciplinary,de2014natural,Kreutzman2014Dasatinib,Yang2012Anti,Liu2013BRAF,Alleneaak9679Combined,Hanahan2000Hakkmarks,
Hanahan2011Hakkmarks,Fouad2011Revisiting,Wang2016Combination,Weinstein2006Mechanism,Pagliarini2015Oncogene,Hijikata2018Phase,Mahadevan2014Misc,
Torti2011Oncogene,Fischer2017Census,Guest2016Functional,Deraedt2011Exploiting,Lasalvia-Prisco2012Addition,bozic2013evolutionary,
sneddon2007location, yang2008nf1, hurwitz2004bevacizumab, lim2017mechanisms,brahmer2012safety,rak1995mutant,cairns2006overcoming,
fan2003combinatorial,feldmann2007blockade,radpour2017tracing,grossenbacher2016natural,Miller2005,ren2014anti,weinstein2008oncogene,
sharma2007oncogene,hu2014combining,gerber2005pharmacology,simpson2016cancer,chaurasiya2018oncolytic,gatzka2018targeted,lin2018advances,
polivka2017advances, allen2017combined,el2016combined,lawson2018oncogenic,bressy2017combining,idema2007addelta24}. Why most cancer does not
respond to conventional therapy is unknown \cite{felsher2003cancer,jain2002sustained, beer2004developmental, shachaf2005tumor,
shachaf2005rehabilitation, luo2009principles}. Most common human cancers are epithelial cancers and they are the least treatable with
conventional therapies. Even when epithelial tumors do initially respond to treatment, eventually the tumors reoccur. Scientists do not know why
targeting any specific gene results in tumor regression.

Although many of the new drugs show a significant clinical response, eventually most tumors do reoccur \cite{felsher2003cancer,
jain2002sustained,beer2004developmental,shachaf2005tumor,shachaf2005rehabilitation}. The inactivation of some oncogene can result in
sustained tumor regression. However, the reactivation of the oncogene results in the rapid restoration of neoplastic properties. Oncogene
inactivation is associated with proliferative arrest, differentiation, and apoptosis.

In some tumors, the cell that was initially transformed by MYC activation and later gives rise to cancer may have stem cell features. As a stem
cell, it can give rise to malignant progenitor cells. Upon oncogene inactivation, this cancer stem cells would then differentiate into normal
appearing and functioning cells, but some of the cells retain their latent stem cell properties. Also, it is possible that MYC inactivation
fails to induce complete eradication of a small subset of cells that have acquired genetic events.

The notion of tumor dormancy has been discussed by oncologists in the context that some types of cancers that were apparently cured by
treatment can reoccur even decades later, such as some sarcomas and breast adenocarcinoma. Other cancers can become dormant upon treatment
with conventional therapies, only to reoccur sometimes years later. Sometimes they can exhibit progression to a transformed more aggressive
state. The immune system has been implicated in the mechanism of dormancy. Immune surveillance mechanisms may hold in check tumor cells in a
steady state. Similarly, other mechanisms that regulate the environment such as inflammation and angiogenesis pathways may also influence
tumorigenesis.

The only way to be certain of the treatment of a cancer is the complete eradication of cancer. The phenomenon of tumor dormancy points to the
lack of insight we have into the mechanisms by which cancers are eliminated upon treatment with a targeted therapy. In the era of targeted
therapeutics, no explanation has been provided for why targeting a specific oncogene or tumor suppressor gene would result in the regression of
a tumor. \cite{felsher2003cancer, jain2002sustained, beer2004developmental, shachaf2005tumor, shachaf2005rehabilitation}.

Carcinogenesis in humans is a complex multistep process. Tumorigenesis is caused by genetic changes that activate oncogenes or inactivate
tumor suppressor genes. However, it takes a dysfunctional tumor microenvironment to raise a tumor. Tumors are dependent on angiogenesis for 
growth. The tumor ecosystem includes many facets of the microenvironment, such as the immune system, the extracellular matrix (including the
cancer stem cell niche), and the inflammation cells. Most cancer tumors are generally untreatable using conventional monotherapies. 

There is a vast literature dedicated to tumor-host interaction \cite{Stepanova1979, deVladar2004, Lefever1979, Kuznetsov1994, Kuznetsov1996,
dOnofrio2005, Eftimie2011, Dingli2006, Albano2006, dOnofrio2010,
Tham2008, Donofrio2006, Donofrio2008, Lo2007, Donofrio2007, Ledzewicz2012, Demidenko2006, Caravagna2010, Albano2011, Cattani2008,
Ledzewicz2013, Wilkie2013, Lithgow2006c, Menchon2007, Kozusko2007, dOnofrio2012, ochab2006coexistence, Ledzewicz2011,
Jimenez2011, Albano2012, dOnofrio2013, Schattler2016, Iwami2012, Roesch2014,Ledzewicz2013-2, Caravagna2012, Sahoo2010, Youshan2007,
Frascoli2014, Moummou2014, Ren2012, Isaeva2009, Patanarapeelert2011, Ledzewicz2014, Albano2007, Ledzewicz2014-2, Ledzewicz2015, Spina2014,
Ledzewicz2014-3, dePillis2014, RomanRoman2012, Albano2008, Chis2009, Ledzewicz2010, Brutovsky2008, Ramos2013, Arshad2016, Albano2015,
Ledzewicz2012-2, Gutierrez2012, Tabatabai2013, Bozkurt2014, Ochab2005, Cattani2013, Cattani2010, Anderson2015, Malinzi2014, Cattani2010-2,
Schattler2012, Vadasz2009, Rozova2016, Schattler2015, Bozkurt2015, Tao2010, Chakraborty2018, Dong2014, Spina2015, Sharifi2018,
Cattani2015, Dunia2011-1, Rosch2016, Oana2013, Amini2015, Dunia2011-2, chakraborty2010stochastic, Ivanov2016, Dunia2011-3,
hu2018dynamics, Ochab2005-2, Dunia2011-4, Yousefi2016, Schattler2015cancer,Rezaei2013optimal,wei2017periodically,lo2007functional,
chis2009analysis,Isaeva2010,Ferrer2016modelizacion,Kiakou2014mathematical,Moral2016simulacion,Sbeity2015review,Albano2012therapy,
sahoo2011stochastic,Lo2009,Shakeri2018adaptive,barrea2017optimal,tsallis2009introduction,cabella2011data,barberis2012describing,galvao2008computational,
barberis2015joint,menchon2008regularization,daSilva2013,khordad2016modeling,grau2007non,sotolongo2003behavior}. The dynamical system that we
investigate includes this process. We will discuss this interaction in detail in the following section.

\section{Entropy and the Model}
\label{Sec:Entropy}

In Ref.~\cite{Calderon1991} a physical justification for the Gompertz's law of tumor growth is presented. The deduction is based on
the concept of entropy. The entropy equation used in Ref.~\cite{Calderon1991} is the well-known Boltzmann-Gibbs extensive entropy. González
\emph{et al.} have used the new non-extensive entropy \cite{tsallis1988possible, curado1991generalized, tsallis1995statistical,
tsallis1998role, botet2001phase, tsallis2009introduction} in the derivation of a new very general evolution equation for tumors
\cite{gonzalez2006}. The logistic, Gompertz, exponential and power laws are particular cases of the new equation. However, it includes many
other cases. The non-extensive parameter $q$ \cite{tsallis1988possible, curado1991generalized, tsallis1995statistical,
tsallis1998role, botet2001phase, tsallis2009introduction, gonzalez2006} plays an important role in the new model. Different types of tumors
possess different values of the non-extensive parameter $q$.

Boltzmann entropy is appropriate if the microscopic interactions are short-ranged, the effective microscopic memory is short-ranged, and/or
the boundary conditions are nonfractal. Non-extensive entropy has been developed for systems with long-range interactions, long-range
microscopic memory, and systems which possess fractal or multifractal properties \cite{tsallis1988possible, curado1991generalized,
tsallis1995statistical, tsallis1998role, botet2001phase, tsallis2009introduction, gonzalez2006}.

The new generalized equation for untreated tumor growth is the following
\begin{equation}
 \label{Eq01}
 \frac{dX}{dt}=\frac{KX_{\infty}}{q-1}\left[1-\left(\frac{X}{X_{\infty}}\right)^q-\left(1-\frac{X}{X_{\infty}}\right)^q\right],
\end{equation}
where $K$ is a certain free parameter and $X_{\infty}$ is the asymptotic value of $X(t)$ when $t\to\infty$. All the published experimental data
that we know can be described by this new equation \cite{gonzalez2006}.

In the present paper, we will investigate the following dynamical system
\begin{eqnarray}
 \label{Eq02}
 \frac{dX}{dt}=\frac{KX_{\infty}}{q-1}\left[1-\left(\frac{X}{X_{\infty}}\right)^q-\left(1-\frac{X}{X_{\infty}}\right)^q\right]-RX-bXY-C(t)X,\\
\label{Eq03}
 \frac{dY}{dt}=d\left(X-eX^2\right)Y-fY+V,
\end{eqnarray}
where $X$ denotes the tumor volume and $Y$ the antibodies density. Equation (\ref{Eq02}) describes the reproduction of the cancer cells,
which are destroyed when they meet the agents of the host response system (term $-bXY$). The reproduction of the agents of the response
system is described by the term $d\left(X-eX^2\right)$, where we consider that, initially, the presence of cancer cells stimulates the
reproduction of $Y(t)$. When the number of cancer cells is large, the reproduction of the antibodies is inhibited, which is why for large
quantities of cancer cells the human defenses are depressed. The term $-fY$ corresponds to the natural death of the antibodies. The term
$V$ represents the external flow of antibodies. The term $-RX$ stands for the natural death of the cancer cells. In most cases, $R=0$.
The term $-C(t)X$ stands for the cell-killing process due to different therapies.

\section{Investigation of the model}
\label{Sec:Investigation}

First, we will consider the case where $q=2$, $X_{\infty}\to\infty$, $R=0$, and $C(t)=0$. Let us redefine $a:=2K$. The dynamical system
(\ref{Eq02}-\ref{Eq03}) can have, in principle, three fixed points
\begin{eqnarray}
 \label{Eq04}
 P_{I}:=\left(X_1,\,Y_1\right):=\left(0,\,\frac{V}{f}\right),\\
 \label{Eq05}
P_{II}:=\left(X_2,\,Y_2\right):=\left(\frac{1}{2e}+\sqrt{\frac{1}{4e^2}-h},\,\frac{a}{b}\right),\\
 \label{Eq06}
P_{III}:=\left(X_3,\,Y_3\right):=\left(\frac{1}{2e}-\sqrt{\frac{1}{4e^2}-h},\,\frac{a}{b}\right),
\end{eqnarray}
where $h:=(fa/b-V)b/ead$. Of course, the fixed points $P_{II}$ and $P_{III}$ exist only when they are real and non-negative. The conditions for
the existence of fixed points $P_{II}$ and $P_{III}$ are the following inequalities
\begin{eqnarray}
 \label{Eq07}
 \frac{1}{4e^2}-h>0,\\
 \label{Eq08}
 h>0.
\end{eqnarray}
The inequality (\ref{Eq08}) is necessary for fixed point $P_{III}$.

We discuss here some results of the dynamical system investigation. If $af\leq Vb$ the fixed point $P_{I}$ is a stable node and the fixed point
$P_{II}$ is a saddle. If $af>Vb$, and $h-1/4e^2<0$, the three fixed points exist and are non-negative. Both fixed points $P_{I}$ and
$P_{II}$ are now saddles. Between these two points, there is the point $P_{III}$ which is stable. If $af>Vb$, and $h-1/4e^2>0$,
then there is only one fixed point ($P_I$) which is unstable now. As a result of this, most trajectories tend to infinity. This means the tumor
will grow limitless.

It is helpful to know the eigenvalues of the Jacobian matrix corresponding to the fixed points. For the point $P_{I}$, the eigenvalues are
\begin{eqnarray}
 \label{Eq13}
 \lambda^{(I)}_1=a-\frac{Vb}{f},\\
 \label{Eq14}
 \lambda^{(I)}_2=-f.
\end{eqnarray}

\begin{figure}
\begin{center}
\scalebox{0.3}{\includegraphics{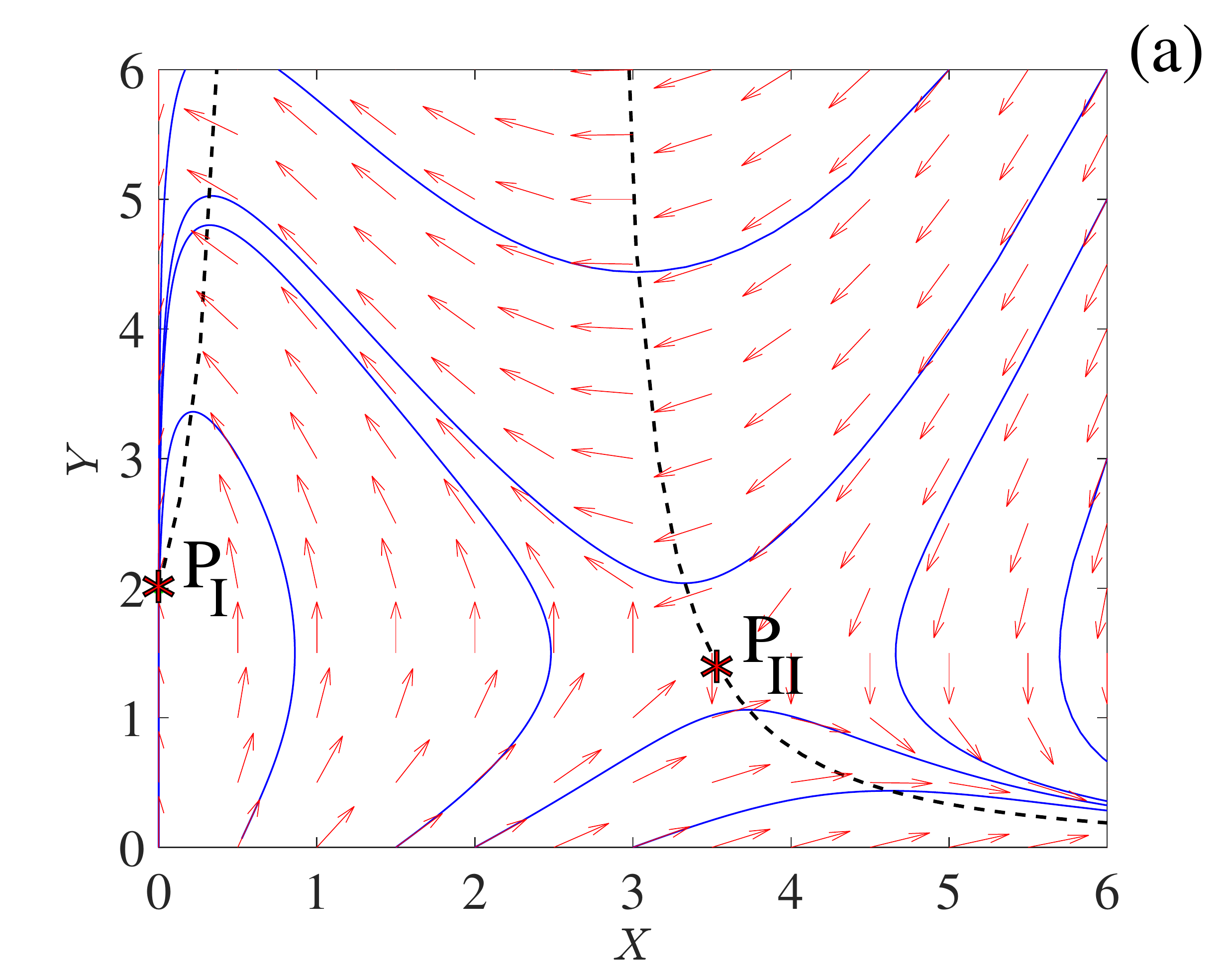}}\scalebox{0.3}{\includegraphics{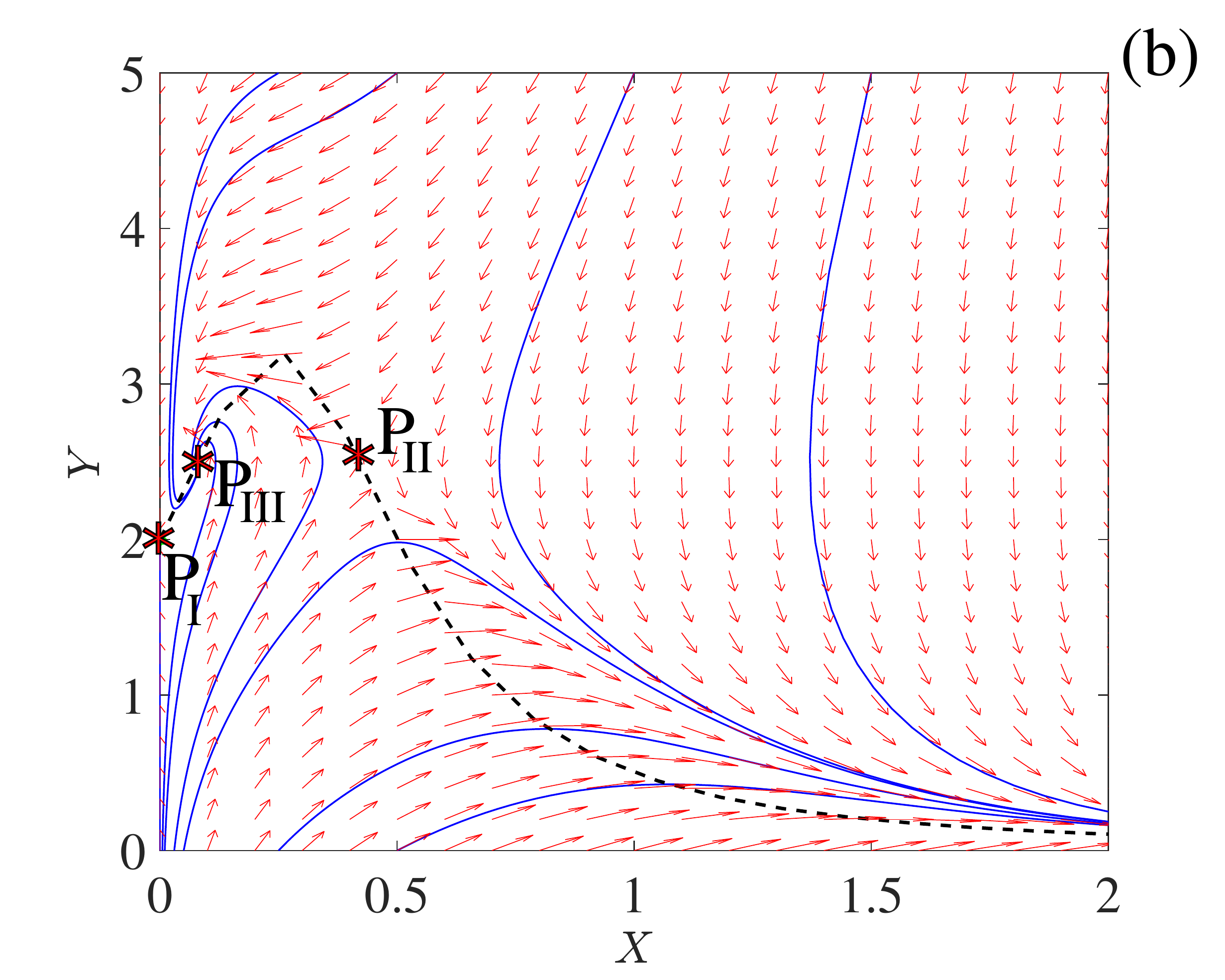}}
\scalebox{0.3}{\includegraphics{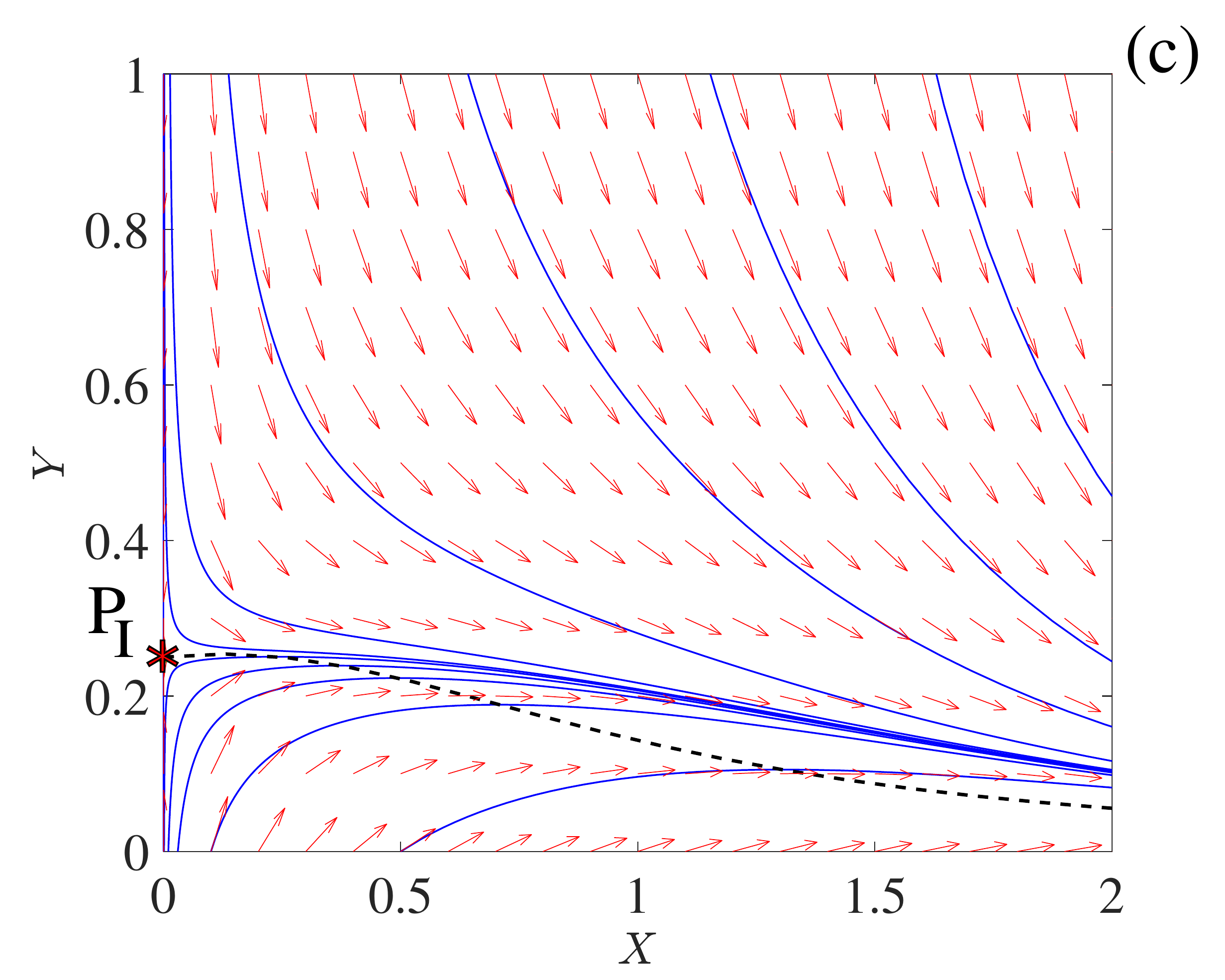}}\scalebox{0.3}{\includegraphics{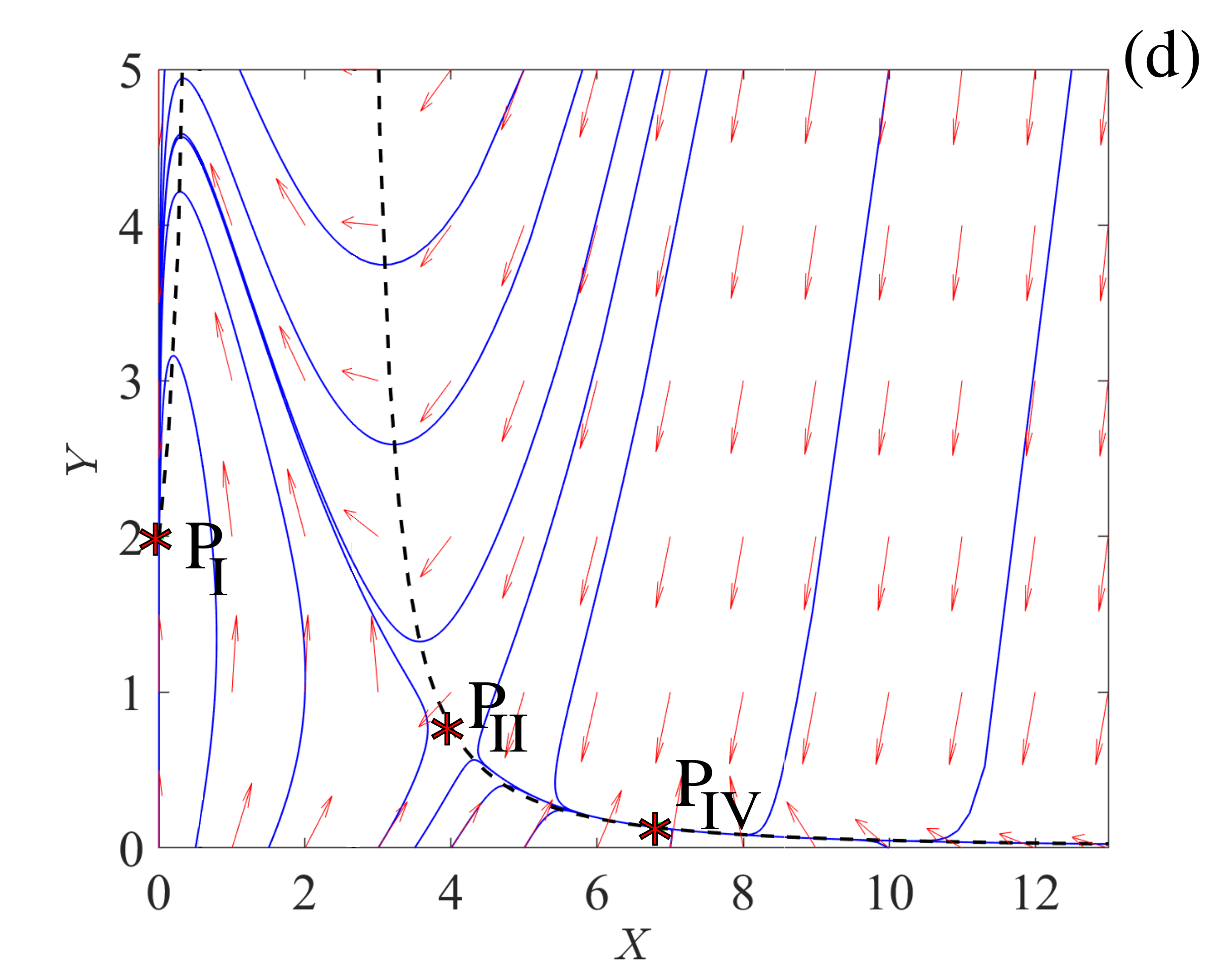}}
\scalebox{0.3}{\includegraphics{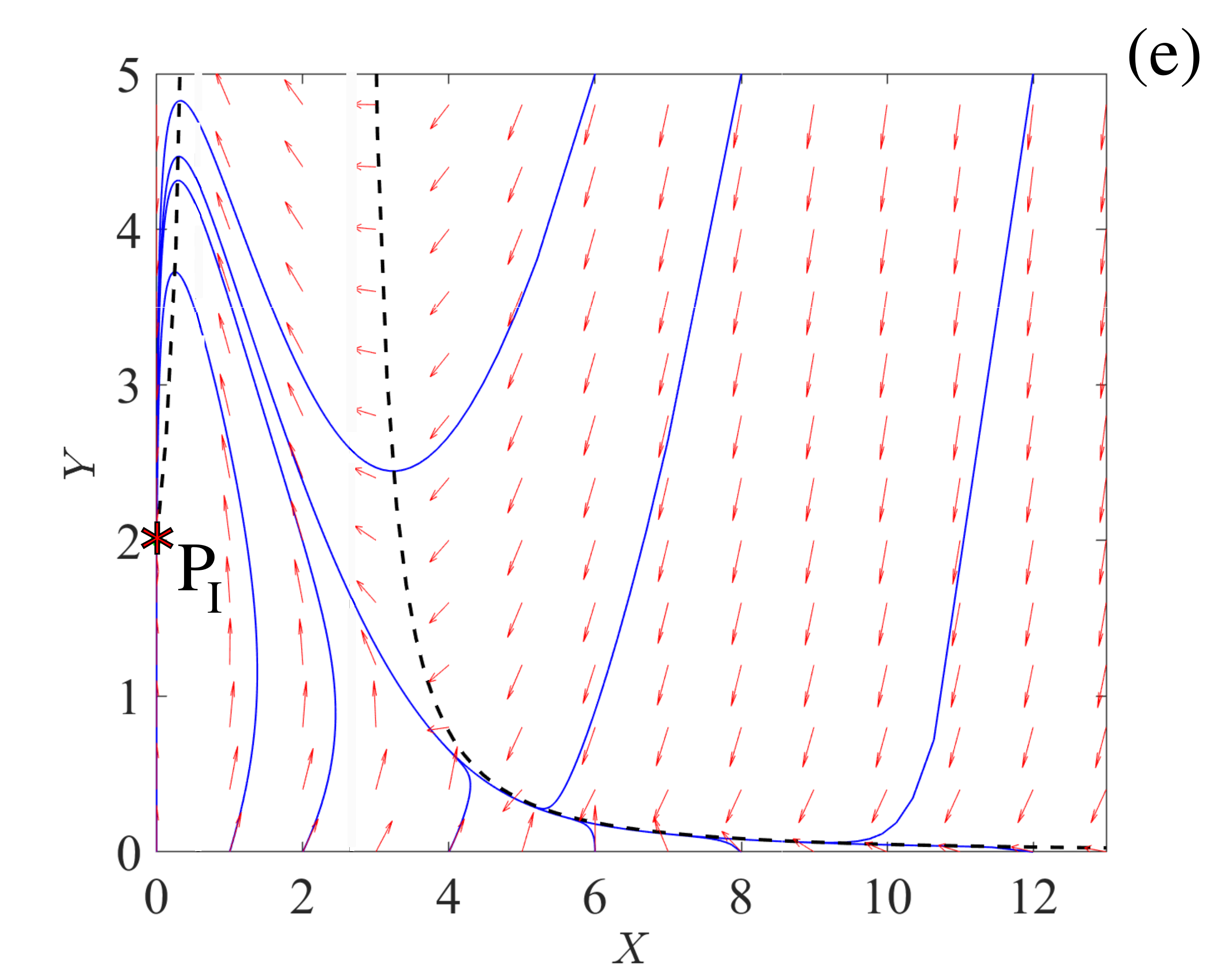}}
\end{center}
 \caption{Phase-space portraits corresponding to the behavior of dynamical system (\ref{Eq02}--\ref{Eq03}). Isoclines are shown in dashed lines.
 See the detailed discussion in the  main text. \textbf{(a)} There are two fixed points: $P_{I}$ is a stable node, $P_{II}$ is a saddle point. The phase trajectories on the left of
 the separatrix lead to the fixed point $P_{I}$. The phase trajectories on the right of the separatrix lead to a fatal outcome. \textbf{(b)} There
 are three fixed points. The point $P_I$ is now a saddle point, which is always unstable. The new point $P_{III}$ is stable, which leads to
 dormancy. \textbf{(c)} There is only one fixed point: Point $P_I$ is a saddle point. Almost all trajectories lead to a fatal outcome.
 \textbf{(d)} When $X_{\infty}<\infty$ (but very large), there is an additional stable fixed point that usually corresponds to a large value of
 $X$. All the trajectories on the left of the separatrix lead to this point. \textbf{(e)} When $X_{\infty}$ is small (this is related to the
 microenvironment, among other factors), it is possible that there is only a stable fixed point (point $P_I$ where $X=0$). All the trajectories
 lead to this point. This is a very favorable situation. \label{Fig01Rep}}
\end{figure}

For the point $P_{II}$, the eigenvalues are
\begin{eqnarray}
 \label{Eq15}
 \lambda^{(II)}_1=\frac{1}{2}\left[B+\sqrt{B^2-4A}\right],\\
 \label{Eq16}
 \lambda^{(II)}_2=\frac{1}{2}\left[B-\sqrt{B^2-4A}\right],
\end{eqnarray}
where $B:=(X_2-eX_2^2)d-f$, $A:=bd(1-eX_2)X_2Y_2$. For the point $P_{III}$, the eigenvalues are
\begin{eqnarray}
 \label{Eq17}
 \lambda^{(III)}_1=\frac{1}{2}\left[G+\sqrt{G^2-4H}\right],\\
 \label{Eq18}
 \lambda^{(III)}_2=\frac{1}{2}\left[G-\sqrt{G^2-4H}\right],
\end{eqnarray}
where $G=d(X_3-eX_3^2)-f$, $H=bd(1-eX_3)X_3Y_3$. In the neighborhood of point $P_{II}$, the separatrix can be approximated by the straight line
\begin{equation}
 \label{Eq19}
 Y=-\left(\frac{\lambda^{(II)}_2}{bX_2}\right)X+\frac{a+\lambda^{(II)}_2}{b}.
\end{equation}

Any point corresponding to initial conditions of the Cauchy problem on the right of the separatrix leads to a fatal outcome. On
the other hand, if the initial conditions correspond to a point located on the left of the separatrix, the system will evolve to a stable fixed point. See
the general dynamics in Fig.~\ref{Fig01Rep}. Using Eq.~(\ref{Eq19}) we can calculate approximately the threshold or critical tumor volume that would
increase limitlessly when the number of antibodies is zero:
\begin{equation}
 \label{Eq20}
 X_{\mbox{crit}}=\left(1+\frac{a}{\lambda^{(II)}_2}\right)X_2.
\end{equation}

\begin{figure}
\begin{center}
\scalebox{0.33}{\includegraphics{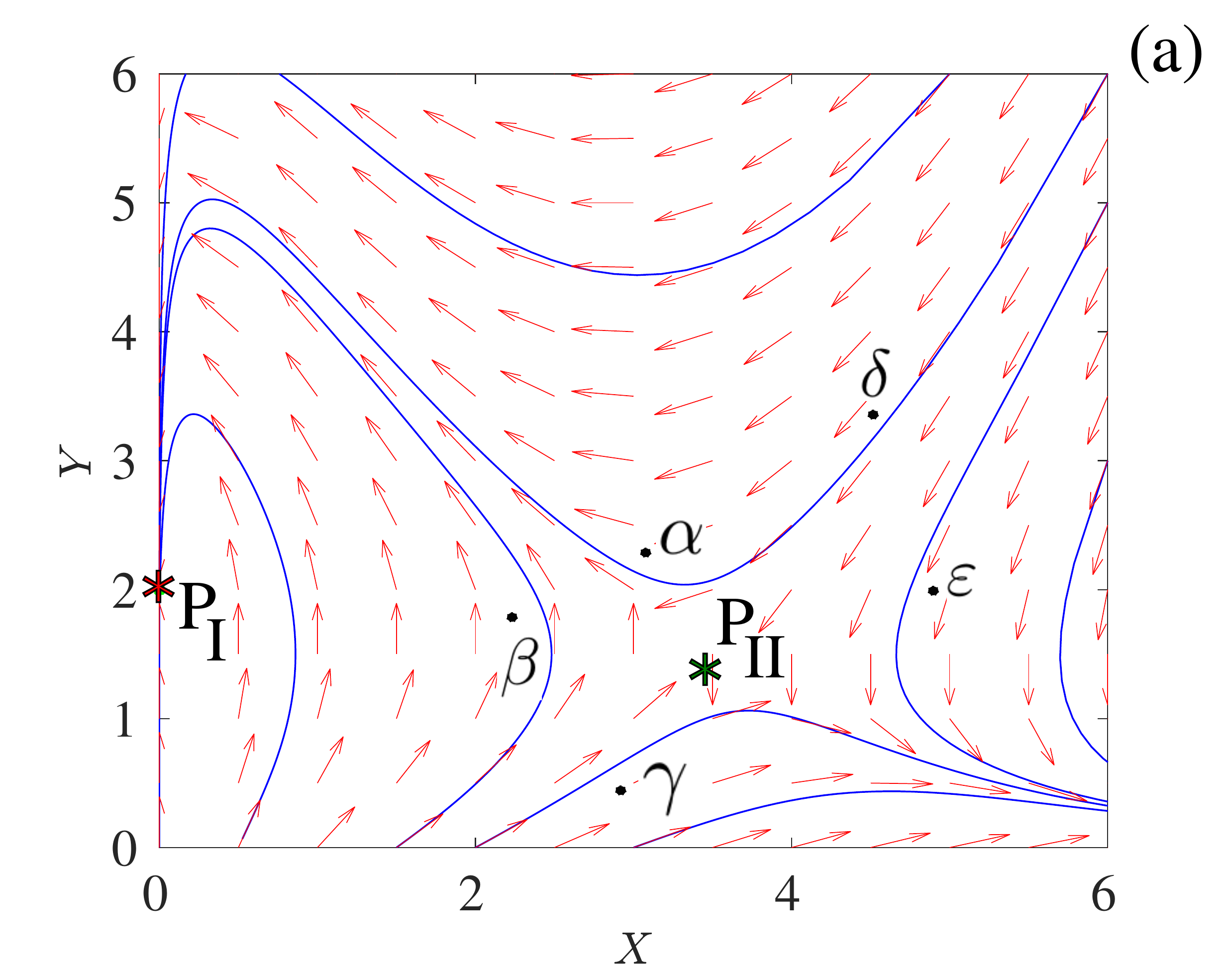}}\scalebox{0.33}{\includegraphics{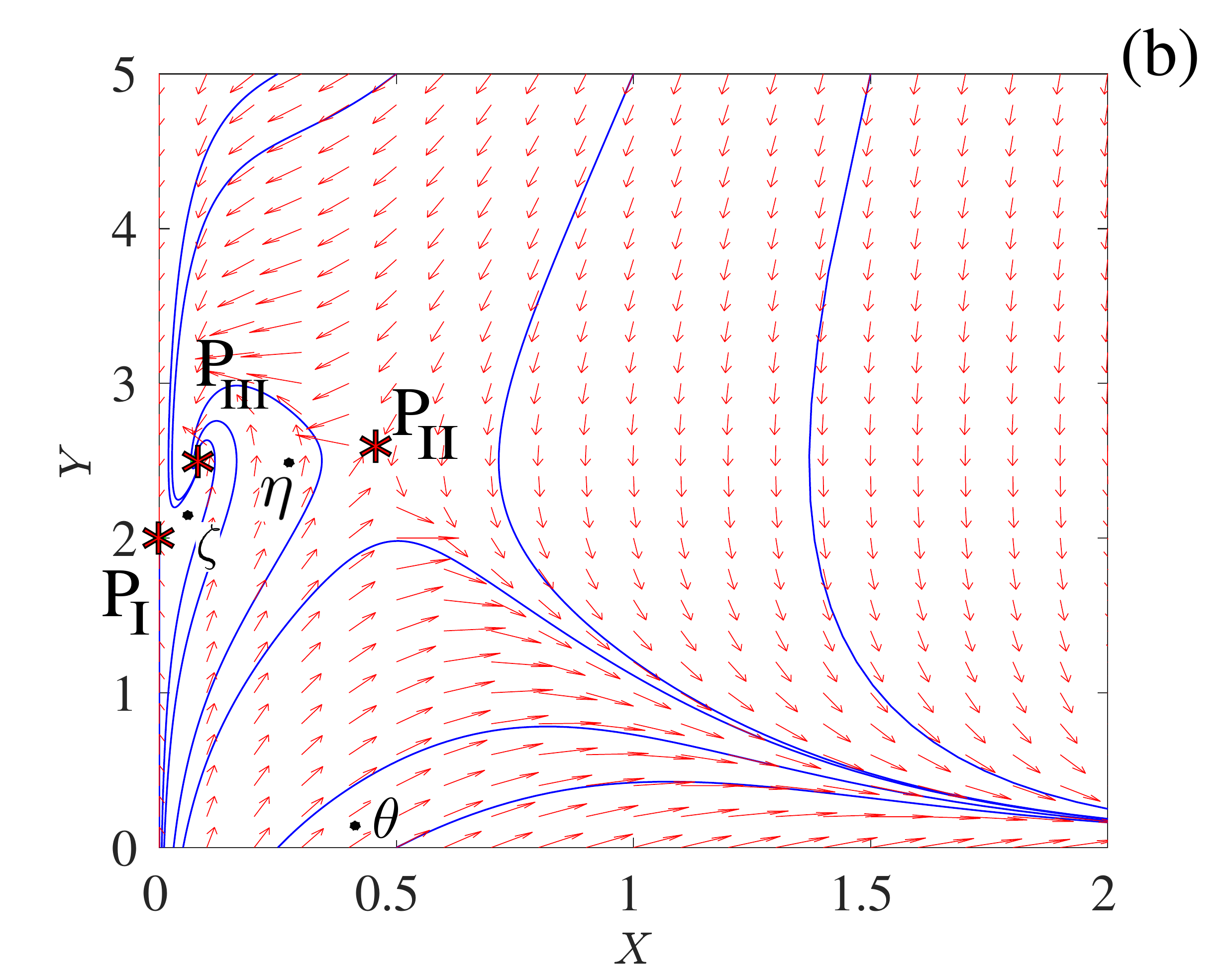}}
\scalebox{0.33}{\includegraphics{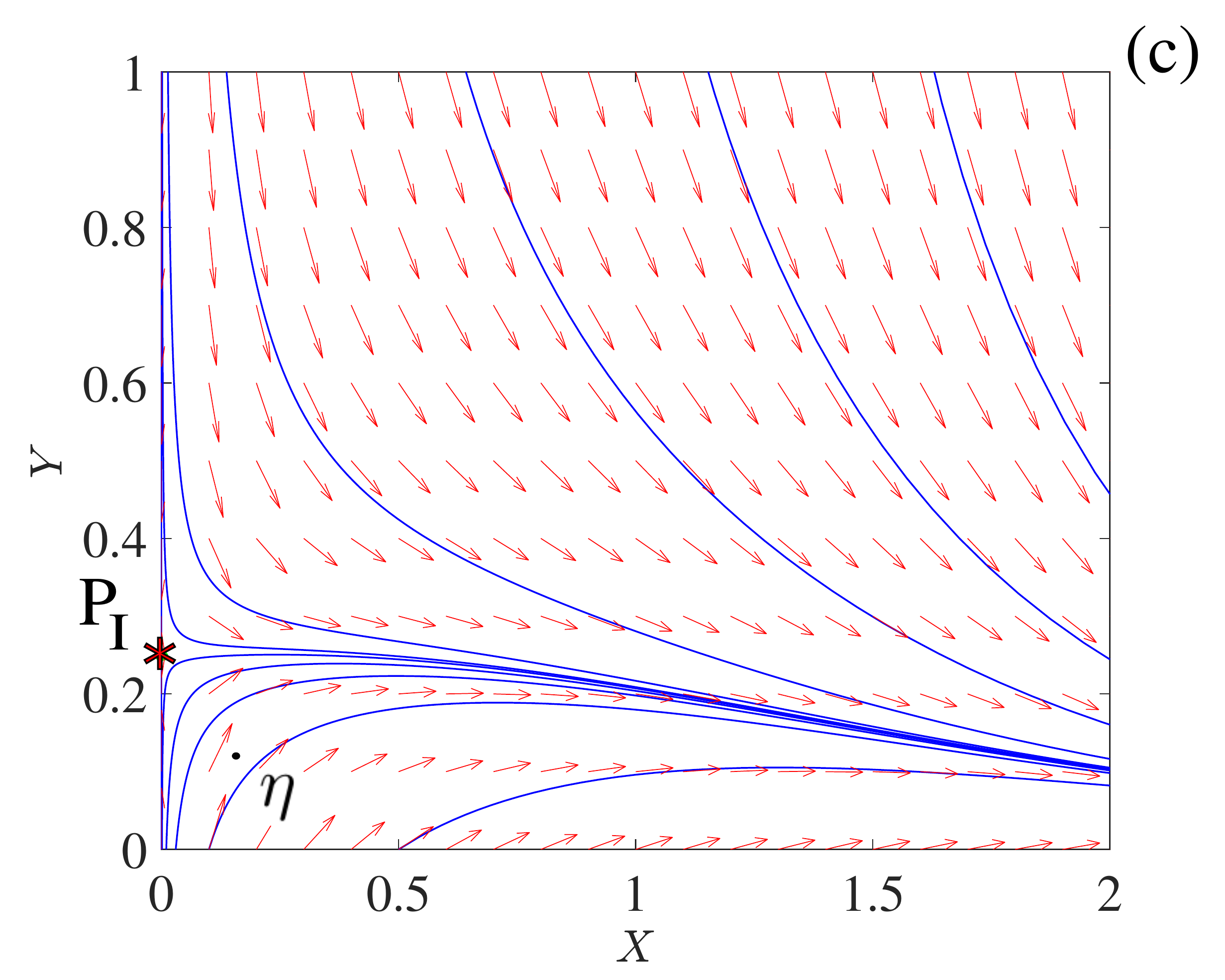}}\scalebox{0.33}{\includegraphics{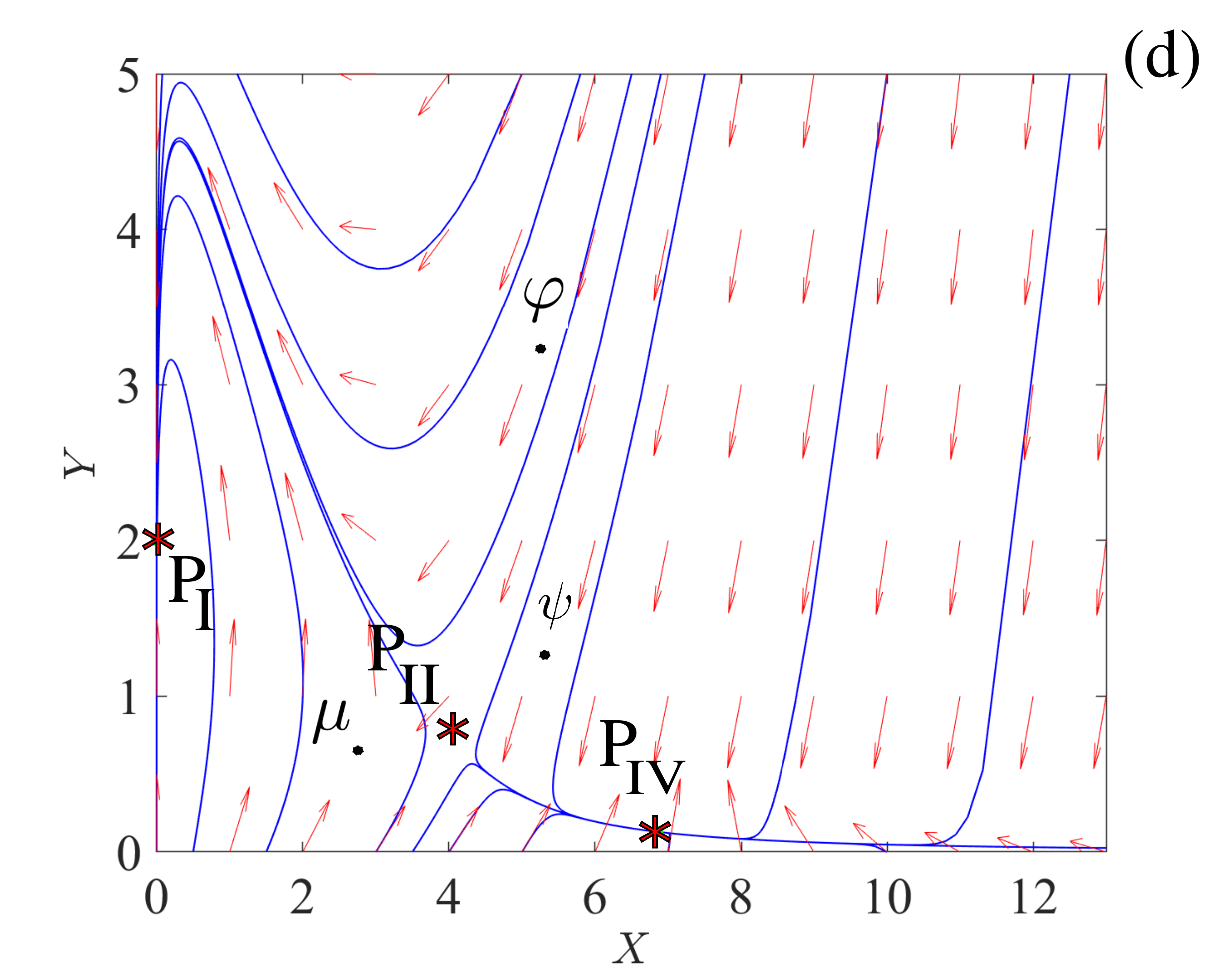}}
\scalebox{0.33}{\includegraphics{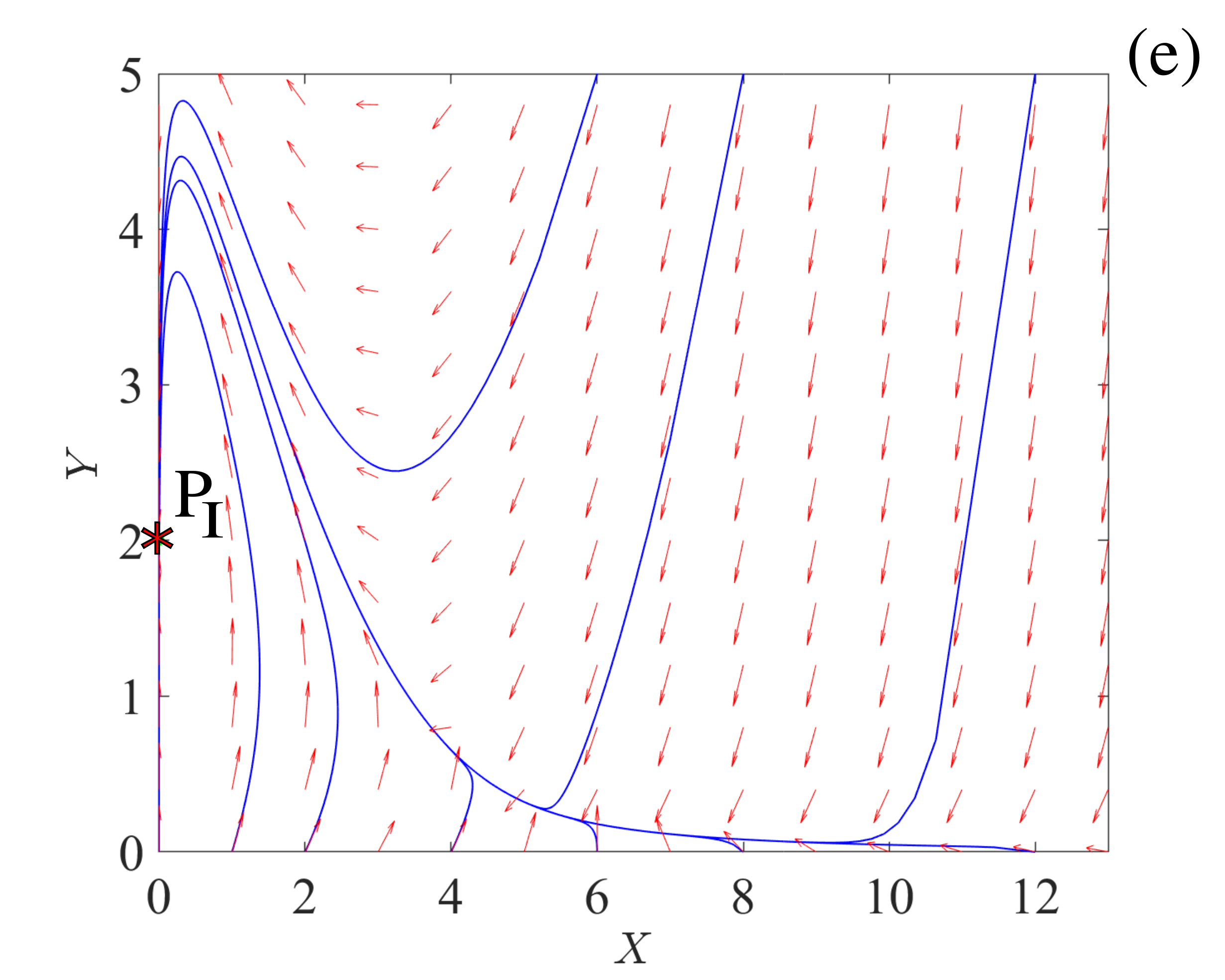}}
\end{center}
 \caption{The fate of different initial conditions. \textbf{(a)} Points $\alpha$, $\beta$, and $\delta$ will ride phase trajectories that tend to the
 fixed point $P_I$ where $X=0$. Note that for point $\delta$ the tumor size is large. Nevertheless, the tumor size will be reduced to zero. On the
 other hand, points $\gamma$ and $\varepsilon$ will ride phase trajectories that lead to a fatal outcome. \textbf{(b)} The separatrix of saddle
 point $P_I$ is very close to the axis $Y$. Initial condition $\eta$ will tend to fixed point $P_{III}$. Meanwhile, the initial condition
 $\theta$ conducts to a fatal outcome. \textbf{(c)} Almost all initial conditions tend to a fatal outcome. \textbf{(d)} Initial conditions
 $\varphi$ and $\mu$ tend to the ``good'' point $P_I$ where $X=0$. Whereas the initial condition $\psi$ tends to a fatal outcome. \textbf{(e)}
 The limit of all initial conditions is $X=0$. \label{Fig01}}
\end{figure}

Note that the different phase space topologies represented in Fig.~\ref{Fig01Rep} depend on the competition between the host-tumor ``forces''.
Fig.~\ref{Fig01}a shows a strong host defense in principle able to reduce to zero any small tumor. Fig.~\ref{Fig01Rep}b shows an
equilibrium between the host defenses and the tumor strength. This equilibrium can conduct to the formation of stable cancer structures. They are
controlled for now. However, they are very dangerous. After some time, if the immune system is depressed for some reason, these structures can
expand. Finally, Fig.~\ref{Fig01Rep}c shows a case where the host defenses cannot sustain the struggle with cancer.

Other methods can also be useful to analyze the dynamics. The isoclines of the system can be obtained from the expression
\begin{equation}
 \label{Eq21}
 \frac{dY}{dX}=\frac{d(X-eX^2)Y-fY+V}{aX-bXY}:=m,
\end{equation}
where $m$ is the slope of the phase trajectory. The explicit expression for the isoclines is
\begin{equation}
 \label{Eq22}
 Y=\frac{amX-V}{d(X-eX^2)+bmX-f}.
\end{equation}

Let us analyze the main isoclines. For $m=\infty$, we have two lines, namely $X=0$ and $Y=a/b$. For $m=0$, we have the isocline
$Y=V/(f-dX-deX^2)$. We can observe the isoclines in dashed lines in Fig. \ref{Fig01Rep}. It is very important to study the different cases
$d>4ef$ and $d<4ef$. A careful analysis of the behavior of the phase trajectories allows us to conclude that the condition
\begin{equation}
 \label{Eq26}
 d>4ef,
\end{equation}
is favorable for the patient. Although it is not a necessary condition, it is a sufficient condition to avoid the dynamics shown in
Fig.~\ref{Fig01}c. Be aware that the zone
\begin{equation}
 \label{Eq27}
 Y<\frac{a}{b},\quad X>\frac{1}{2de}\left(d-\sqrt{d^2-4def}\right),
\end{equation}
is still very dangerous.

In many cases, it is convenient to re-write the system (\ref{Eq02}--\ref{Eq03}) as one equation where the only unknown is $X(t)$ and $Y$
disappears:
\begin{equation}
 \label{Eq29}
 \frac{d^2X}{dt^2}+\left[f-d\left(X-eX^2\right)\right]\frac{dX}{dt}
 -\frac{1}{X}\left(\frac{dX}{dt}\right)^2=(af-Vb)X-adX^2+adeX^3.
\end{equation}

Eq.~(\ref{Eq29}) can be written in the form 
\begin{equation}
 \label{Eq30}
 \frac{d^2X}{dt^2}+F_{\mbox{dis}}\left(X,\,\frac{dX}{dt}\right)=-\frac{dU(X)}{dX},
\end{equation}
where
\begin{equation}
 \label{Eq31}
 U(X)=\frac{1}{2}(Vb-af)X^2+\frac{1}{3}adX^3-\frac{1}{4}adeX^4,
\end{equation}
and 
\begin{equation}
 \label{Eq32}
 F_{\mbox{dis}}=\left[f-d\left(X-eX^2\right)\right]\frac{dX}{dt}
 -\frac{1}{X}\left(\frac{dX}{dt}\right)^2.
\end{equation}

\begin{figure}
\begin{center}
\scalebox{0.35}{\includegraphics{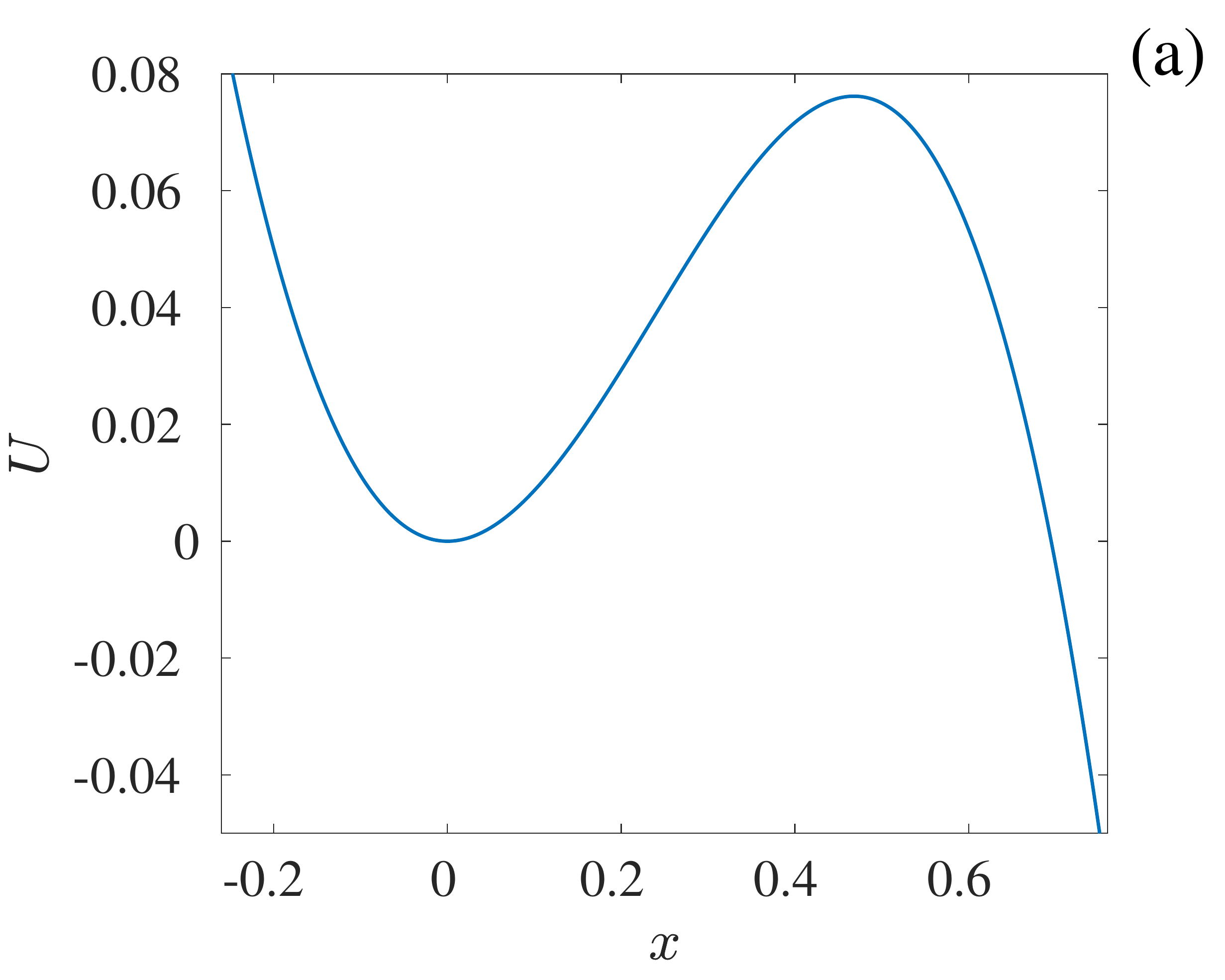}}\scalebox{0.35}{\includegraphics{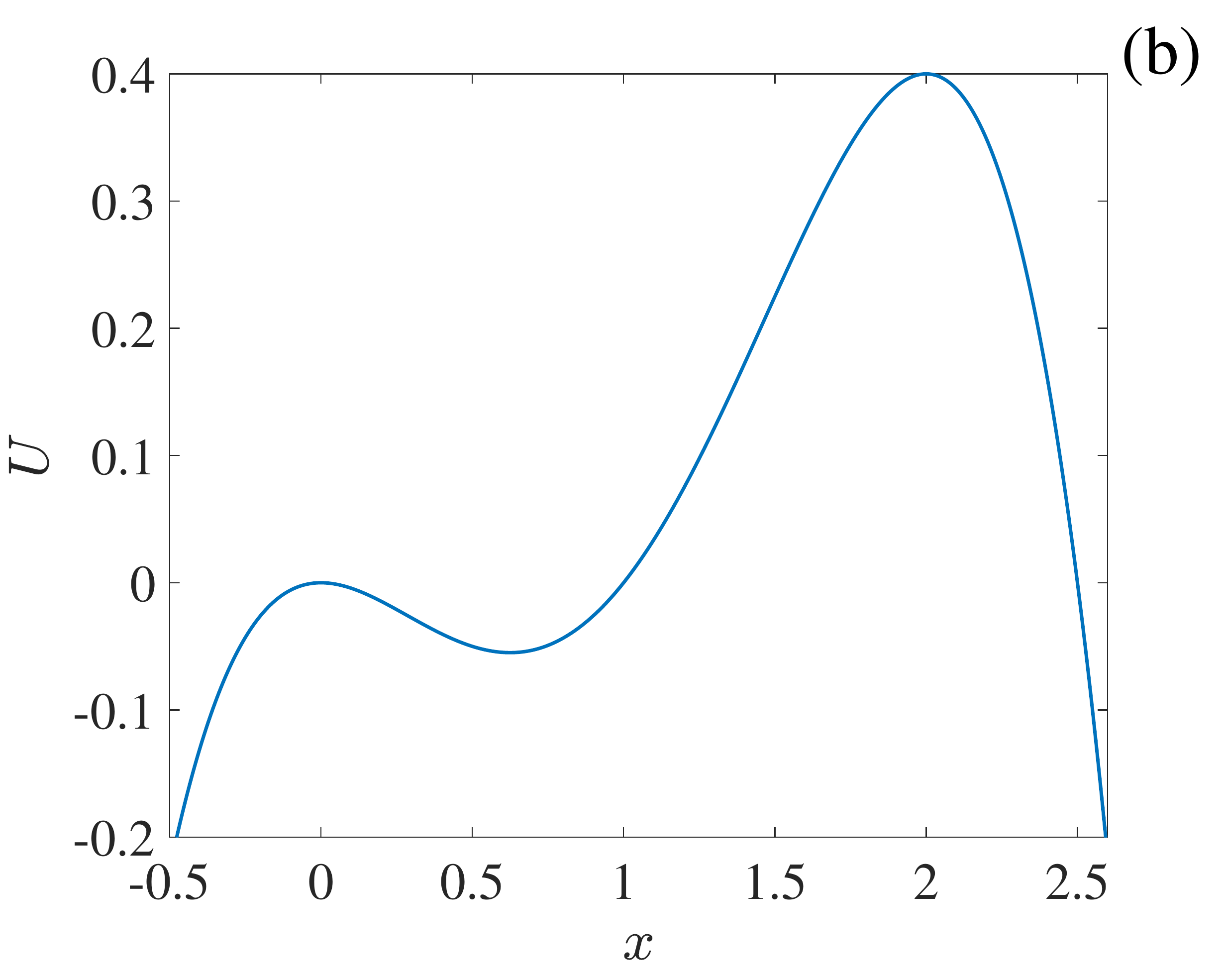}}
\scalebox{0.35}{\includegraphics{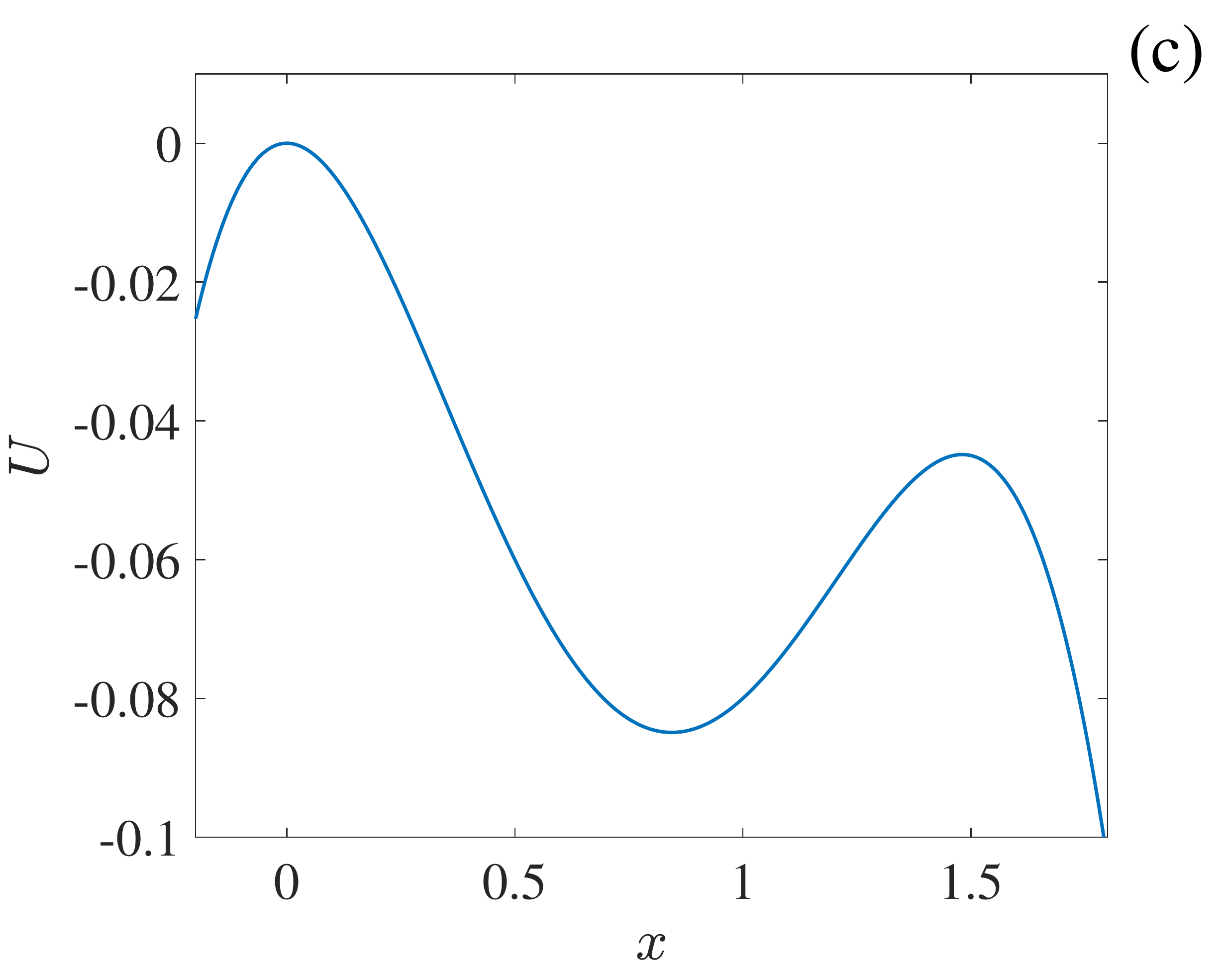}}\scalebox{0.35}{\includegraphics{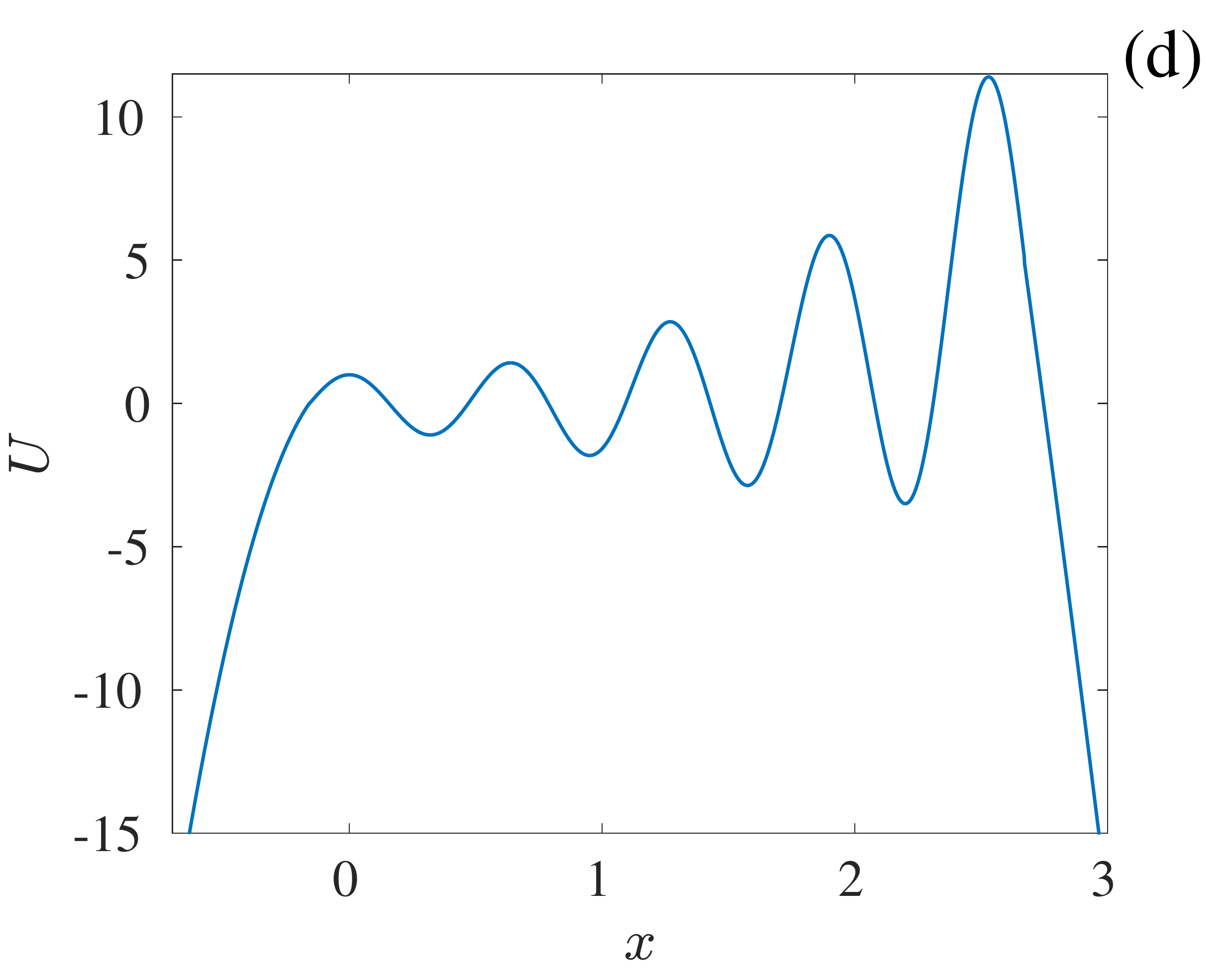}}
\end{center}
 \caption{Possible potentials $U(X)$ for Eq.~(\ref{Eq30}). \textbf{(a)} Case when fixed point $P_I$ is a stable node (it corresponds to the local
 minimum of the potential). Additionally, there is a saddle point (the maximum). \textbf{(b)} Case when there are three fixed points. Fixed point
 $P_I$ is a saddle, which is unstable. The ``jump'' over the potential barrier on the right is harder than in the case shown in Fig.~\ref{Fig02}c.
 \textbf{(c)} The same potential represented in Fig.~\ref{Fig02}b with different parameters. The height of the potential barrier on the right
 is lower than in the case of Fig.~\ref{Fig02}b. This case is a more dangerous scenario for the patient. \textbf{(d)} The Eq.~(\ref{Eq30}) can
 be a more generalized model with a more general potential $U(X)$ with many minima and maxima. The Fig.~\ref{Fig02}d shows an arbitrary example.
 \label{Fig02}}
\end{figure}

Equation (\ref{Eq30}) and the potentials shown in figure \ref{Fig02} suggest that resonance could play an important role in cancer development.
In these pictures, we can observe the existence of potential wells and barriers. The fictitious particle can be trapped in one of those potential
wells. In this case, we can have tumor dormancy. However, external perturbations or the change in the potential $U(X)$ due to aging or other
diseases can lead to the decrease in height of the right side potential barrier generating new cell proliferation that can provoke the creation
of large cancer tumors.

The model can be generalized using Eq.~(\ref{Eq30}) and a more general potential $U(X)$ with many maxima and minima. An example of a possible new 
potential is shown in Fig.~\ref{Fig02}d. The creation of a higher barrier and a lower potential well can be related to the targeting of an
oncogene. However, this will lead to dormancy. This is not a cure. This phenomenon is in agreement with experiments. Later, a change in the 
potential due to some events discussed in the text can lead to regrowth.

When the ``fictitious'' particle is oscillating inside one of the potential wells, an external oscillating perturbation can produce a resonant
behavior of the particle motion leading to an increase of the amplitude that can make the particle to jump over the potential barrier and the
tumor size to increase. Moreover, the formation of a potential barrier can be related to the inactivation of some oncogene. This process can stop
cancer development for a while. This process is related to oncogene addiction. This idea could solve one of the most intriguing problems in
recent cancer research. The external perturbation that can create resonance could be a recurrent lesion, injury or tissue damage. For example,
this may be generated by smoking or frequent radiation exposure.

Equation (\ref{Eq30}) can be considered as Newton's equation for a particle moving in the potential $U(X)$ in the presence of dissipative
forces. The potential $U(X)$ that corresponds to the situation shown in Fig.~\ref{Fig01}a has the shape depicted in Fig.~\ref{Fig02}a. Notice
that when the inequality
\begin{equation}
 \label{Eq33}
 9e(Vb-af)+2ad>0
\end{equation}
is satisfied, the height of the maximum on the right is higher than the maximum at point $X=0$ (see Fig.~\ref{Fig02}b). Otherwise, $U(X)$ has the
shape shown in Fig.~\ref{Fig02}c. The case depicted in Fig. \ref{Fig02}c is more favorable for the particle to jump above the maximum on the
right, which is equivalent to unlimited growth in the number of cancer cells. A careful analysis shows that the condition
\begin{equation}
 \label{Eq34}
 2d>9ef
\end{equation}
is favorable for the patient. Moreover, in order to limit the ``jumps'' above the maximum on the right, it is convenient if the eigenvalues of
fixed point $P_{III}$ are real and negative. In this case, this point behaves as a stable node (and not like a focus). This situation limits the
``inertia'' in the particle motion.

When $X_{\infty}$ is a large but finite number, the dynamical system will have an additional fixed point that will be a stable equilibrium
position. Let us call this equilibrium fixed point $P_{IV}$ (see figures \ref{Fig01Rep}d and \ref{Fig01}d). Those phase trajectories that are on
the right of the separatrix, instead of approaching infinity, will be approaching fixed point $P_{IV}$.

\section{The general model}
\label{Sec:GeneralModel}

For $q>1$, the topological behavior of the dynamical system (\ref{Eq02}--\ref{Eq03}) is very similar to the dynamics explained in Section
\ref{Sec:Investigation}. However, when $q\leq1$, the fixed point $P_I$ will be always unstable. \emph{This means that it is impossible to reduce
the tumor volume to zero}. This is one of the most remarkable results of this investigation. This finding will play a very important part in the
explanation of the phenomena discussed in the Introduction and in the design of new therapies.

\section{Conventional therapies}
\label{Sec:Conventional}

As we have seen in the previous sections, for $q>1$, the parameters of the system and the initial conditions play an important role in the
outcome. The tumor-host interaction is decisive. There are situations where the immune system by itself can reduce to zero the tumor volume.
Under other circumstances, the tumor volume will increase leading to fatal consequences.

If we apply conventional therapy $C(t)=C_o$ in the system (\ref{Eq02}--\ref{Eq03}) (where $C_o$ is a constant), for $q>1$, the cancer cure can be accelerated.
If $q\leq1$, then for any value of $C_o$, $X(t)$ is never reduced to zero. The fixed point $P_I$ is always unstable.

Most cancer tumors are generally untreatable using conventional monotherapies. Some cancers can become dormant upon treatment with conventional
therapies (this can be observed as a stable fixed point with $X>0$) only to reoccur later. The parameters of the system can change due to illness,
age, immune system depression, et cetera. In some cases, the stable finite fixed point and the saddle points can disappear. As the fixed point
$P_I$ is unstable, the dynamics will induce a lethal increase in the tumor volume. We would like to stress again that most cancer tumors are
generally untreatable using conventional monotherapies.

\section{How the model can guide physicians to invent better therapies}
\label{Sec:Physicians}

An extensive study of many experimental, clinical, and theoretical papers (including ours) leads us to conclude that $q$ depends on the genes.
Our analysis in the previous sections shows that for $q\leq1$, the fate of the patient is most certainly lethal.
All this investigation leads to combination therapies \cite{siegel2014cancer, coit2013melanoma, chapman2011improved, hauschild2012dabrafenib,
hodi2010improved, robert2014anti, berger2012melanoma, davies2002mutations, nikolaev2012exome, ribas2011braf, trudel2014clinical, 
trunzer2013pharmacodynamic, flaherty2012combined, prickett2009analysis, guo2011phase, kluger2011phase, schreiber2011cancer, dong2002tumor,
wolchok2013nivolumab, beer2011randomized, rosenberg2011durable, draube2011dendritic, leonhartsberger2012quality, huber2012interdisciplinary,
de2014natural, Kreutzman2014Dasatinib, Yang2012Anti, Liu2013BRAF, Alleneaak9679Combined}. We have to change $q$ first, then we can use therapies
that change the parameters in such a way that the fixed point $P_I$ is asymptotically stable (a stable node), and finally, we need to apply
therapies that will help the phase trajectory to go to the point $P_I$. The only way to have an effective treatment of cancer is the complete
eradication of all the tumors. All this points to a combination of oncogene-targeted therapy, tumor-suppressor gene-targeted therapy,
immunotherapy, anti-angiogenesis therapies, and tumor-cell killing therapies. Other combinations can also be successful.

\subsection{Combinational therapies}
\label{Sec:Combinational}

Most cancer tumors are generally untreatable using conventional monotherapy. However, these tumors can be effectively treated with rationally
designed combinational therapies. Our work is not completed until patients are cured of their disease. Essential to our ability to optimally kill
cancers are approaches that should be combined with therapies that exploit oncogene addition and the exploitation of somatic loss-of-function
tumor suppressor gene mutations. A second very important part of the combination is the use of our emerging ability to reactivate immune
response to tumors \cite{hauschild2012dabrafenib, hodi2010improved}. If we are able to effectively combine these emerging therapeutic options,
the road toward a cure is becoming clear.

We need therapies with the potential to target both tumor cells and tumor microenvironment. Drugs that target oncogenes can be effective in the
treatment of some cancers. However, most tumors do reoccur. We have found that the success of the new therapeutic agents can be seen when used in
combination with other kinds of therapies, including conventional treatments.

Oncogene addiction and cancer stem cells are related to $q$. FLT3 is a receptor tyrosine Kinase class III that is expressed on by early
hematopoietic progenitor cells and plays an important role in hematopoietic stem cell proliferation, differentiation, and survival. The addition
of multitargeted Kinase inhibitor midostaurin to standard chemotherapy significantly prolonged overall and event-free survival among patients
with Acute Myleoid Leukemia and FLT3 mutation \cite{stone2017midostaurin}.

Combination therapies provide a rational strategy to potentiate efficacy and potentially kill the tumors. Host immunity contributes to the
anti-tumor activity of oncogene-targeted inhibitors within a complex network of cytokines and chemokines. Therefore, combining immunotherapy with
oncogene-targeted drugs may be the key to cancer control. Targeted therapies can achieve impressive and rapid tumor remissions, although these
results are eclipsed by the emergence of resistance mechanisms through selection in heterogeneous tumors, limiting clinical response to a
relatively short duration \cite{aris2015combining}. In contrast, immunotherapy can give rise to long-term cancer control by eliciting active
immune effectors that may lead to a curative response, but in a small group of patients.

Only combination will achieve long-term responses. We need to develop more rationalized combination therapies with oncogene-targeted therapies
and immunotherapy. There are several combinations and trials described in Ref.~\cite{aris2015combining}. The combination therapy provided
a superior antitumor response to either single-agent modality alone. This suggests that combining BRAF-targeted therapy with IL-2 could contribute
to increased tumor destruction.

Combination therapies comprised of oncogene-targeted and immunotherapeutic strategies are a promising emerging approach to cancer treatment.
Patients who have rapidly progressing tumors with druggable mutations can benefit from oncogene-targeted drugs first. The change of the $q$ value
can lead to the conditions for the stability of the fixed point where $X=0$. The stabilization of this fixed point can allow the elimination of
important tumor masses in a short time. Sequential immunotherapy administration could cause tumor immune infiltration and therefore, promote
durable immune control of disease dissemination. All these studies can give rise to more personalized medicine for cancer treatment. Many of
these phenotypic traits can be brought about by genetic alterations that involve the gain-of-function mutation, amplification, and/or
overexpression of key oncogenes together with the loss-of-function mutation, deletion, and/or epigenetic silencing of key tumor suppressors.

There is tremendous complexity in the patterns of mutations in tumors of different origin. A key to successful therapy is the identification of
critical, functional points in the oncogenic network whose inhibition will result in the system failure, that is, the cessation of the
tumorigenic state by apoptosis, necrosis, senescence, or differentiation. We believe these critical structures in the oncogenic network are
related to parameter $q$.

The two mainstay treatment options for cancer today (chemotherapy and radiation) are examples of agents that exploit the enhanced sensitivity of
cancer cells to DNA damage. Scientists do not have a clear molecular understanding of why they eventually fail. The goal of cancer therapy is to
target the hallmarks as tumor-specific liabilities, preferably through the combinatorial application of a number of drugs. Thus, we need a
thorough understanding of the nature of these hallmarks. Our understanding is the following. In order to change $q$, we need both
oncogene-targeted and tumor-suppressor-targeted therapies. It is already clear that each of even the best
therapies applied alone eventually fail in the majority of cases \cite{luo2009principles}. A combinatorial series of several different therapies
applied concurrently is needed to eliminate all of the cancer cells in a patient. Semenza \textit{et al.} is using combination therapy in the
treatment of chemoresistant breast cancer \cite{Samanta2014}. Previous
research revealed that triple-negative breast cancer cells show a marked increase in the
activity of many genes known to be controlled by the protein hypoxia-inducible factor (HIF).
HIF enhances the survival of breast cancer stem cells, which are the cancer cells that must be
killed to prevent collapse and metastasis. These are drugs that block HIF from acting. Semenza's team genetically altered the cancer cells to
have less HIF. Thus the cancer stem cells
were no longer protected from death by chemotherapy, demonstrating that HIF was required for the
cancer stem cells to resist the toxic effects of \emph{paclitaxel}. Triple-negative breast cancer
cells were given paclitaxel plus the HIF inhibitor digoxin. Treatment with digoxin and paclitaxel
decreased tumor size by 30 percent more than treatment with paclitaxel alone. The combination also
decreased the number of breast cancer stem cells. Treatment with digoxin plus a different 
chemotherapy drug, gemcitabine, \emph{brought tumor volumes to zero} within three weeks and
prevented the immediate relapse at the end of the treatment. For physicians, how this will be
accomplished remains to be determined. We hope that our model can help solve this problem.

The different therapies participating in the combination should act synergistically in such a way
that suppressor mutation for the first therapy cannot suppress the second therapy and vice versa.
This combinatorial series of different therapies can provide a cure. Experimental design mathematics can be used a posteriori as an inverse
problem in order to obtain the experimental design matrix. This will allow us to obtain the impact in cancer cell killing of
different therapies. The total cancer cell killing impact should be increased logarithmically, or at least using a
late-intensification schedule \cite{deVladar2004, gonzalez2003}. We believe that using our models as a guide it is
possible to design a proper combination of cancer therapies. Through this combination, it is
possible to convert cancer from a death sentence into a curable disease.

Despite encouraging results, Oncolytic Virus monotherapy based exclusively on virus
replication-induced oncolysis often does not demonstrate all the desired qualities,
especially when tested against virus-resistant malignancies \cite{Wheldon1988, Laird1964, Skipper1971, McCredie1965, Norton1977,
bressy2017combining}. Usually, Oncolytic Virotherapies are engineered to express an exogenous transgene with anti-tumor activity and/or
combined with standard treatments like radiotherapy or chemotherapy.
A number of recombinant Oncolytic Virotherapies expressing a transgene for p53 or another p53 family member (p63 or p73) have been engineered
with the goal of generating more potent Oncolytic Virotherapies that function synergistically with \emph{host-immunity} and/or other therapies
to reduce or eliminate tumor burden. Such transgenes have proven effective at improving Oncolytic Virotherapy and the researchers have shown
mechanisms of p53-mediated enhancement of Oncolytic Virotherapy, provided they have optimized p53 transgenes and explored drug Oncolytic
therapies combinatorial treatments.
There is a study with combination Adenovirus $\Delta$23-p3 \footnote{Adenoviral vectors can be administered at different doses} plus
radiotherapy with good results. Experiments and clinical trials data suggest that Oncolytic Virotherapy-encoded p53 can simultaneously
produce anti-cancer activities while assisting, rather than inhibiting, virus replication in cancer cells. One of the major advantages of
using p53-armed Oncolytic therapy is their enhanced oncotoxicity \cite{bressy2017combining}.

Oncolytic Virotherapies aim not only for direct Oncolysis of cancer cells flowing virus replication, but also stimulation of a host's anti-cancer
(innate and adaptive) immune responses.  Various studies have shown that beneficial p53 functions include the promotion of enhanced anti-tumor
immunity, both innate and adaptive. It is believed that p53 transgene expression would augment the anti-tumor immunity to help eliminate the
tumor during Oncolytic Virotherapy treatment. The engineered adenovirus, SG7605-11R-p53, tested in gallbladder cancer cell lines demonstrated
that infection with p53-11R improved the antitumor effect and prolonged survival, compared with control viruses.

Let us discuss the combination of Oncolytic Viruses p53 transgene with radiotherapy \cite{bressy2017combining}. Because p53 has been shown to
enhance the effects of radiotherapy \cite{bressy2017combining}, researchers tested the combination of Oncolytic Virotherapy with radiotherapy
against glioma cancer cells. Ad$\Delta$24-p53 and radiotherapy increase antitumor efficacy compared with every single treatment. This study
highlights that  Ad$\Delta$24-p53 combined with radiotherapy can \emph{eradicate} tumors, which would otherwise escape Oncolytic Virotherapy as
a monotherapy.
Adenoviral vectors were injected at different doses depending on the administration route. The most common route used a virus dose ranging from
$1\times10^8$ to $7.5\times10^{12}$ VPs. Other injection routes were also utilized such as intra-arterial ($7.5\times10^9$ to $7.5\times10^{13}$
VPs). Regarding bladder and lung cancer, the Adenoviral vectors were transmitted by intravesicular instillation ($7.5\times10^{11}$ to
$7.5\times10^{13}$VPs). The majority of patients who received Oncolytic Virotherapy displayed regression of tumor mass or a  transient
stabilization of their disease.

We propose to use a logarithmic function to calculate the dose \cite{gonzalez2003, gonzalez2006}. Replicating Oncolytic Virotherapy-p53
viruses based on other viruses could be more efficient in future trials, because they may spread within the tumors amplifying p53 expression
and increasing oncolysis. Replicating Oncolytic Viruses could improve the anti-tumor response by the lysis of tumor cells to allow the
release of numerous anti-tumor antigens into the tumor microenvironment. These antigens could be processed
and potentially lead to sustainable adaptive immune responses.
Oncolytic Virus-p53 therapy can simultaneously achieve direct oncolysis and anti-tumor immunity (against several tumor-specific antigens,
including mutant p53). More experiments can be conducted \emph{combining} Oncolytic Virotherapy-p53 viruses with chemotherapy. Additionally, 
the use of statins or the inhibitors of these kinases in \emph{combination} with Oncolytic Virotherapy encoding p53 could provide additional
benefits.

\section{Hallmarks of cancer}
\label{Sec:Hallmarks}

\begin{figure}
  \begin{center}
    \scalebox{0.19}{\includegraphics{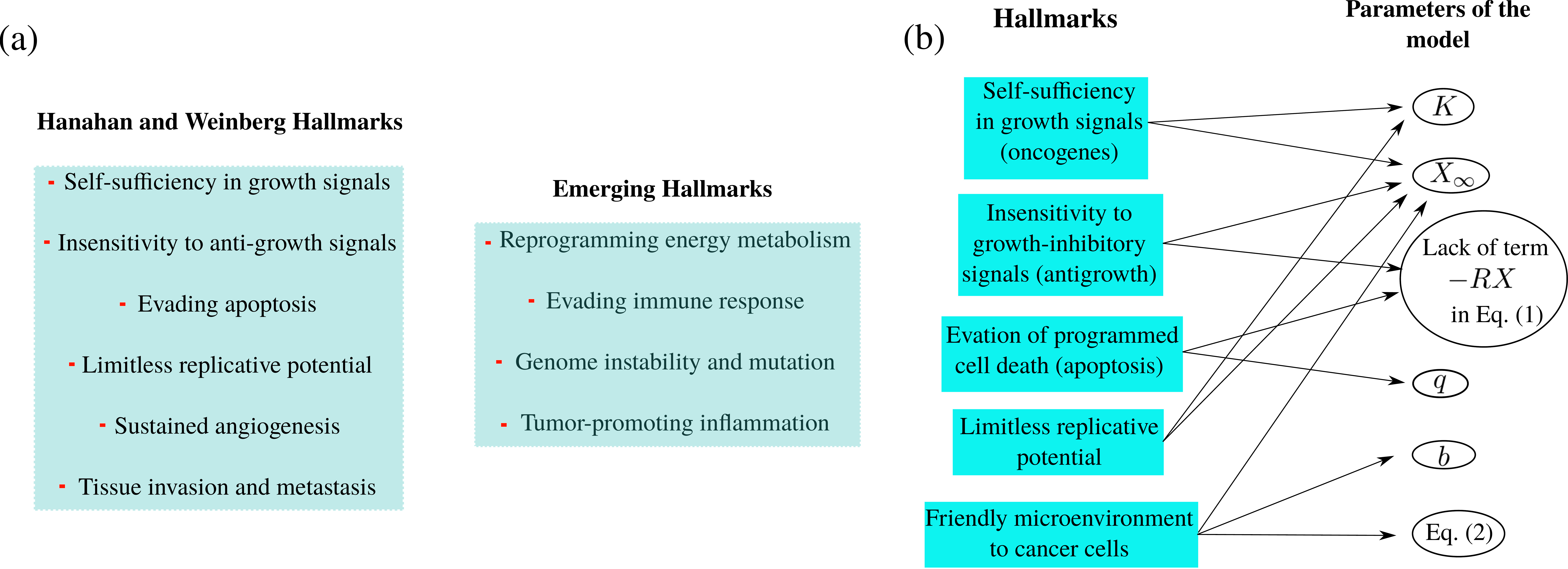}}
  \end{center}
  \caption{\textbf{(a)} Hallmarks of cancer, according to Hanahan and Weinberg \cite{Hanahan2000Hakkmarks}. A hallmark is a distinguishing
  feature of cancer. \textbf{(b)} Connections between the parameters of our model and the hallmarks of cancer.
  \label{Hallmarks}}
\end{figure}

Cancer is a complex collection of distinct genetic diseases united by common hallmarks (see Fig.~\ref{Hallmarks}a) \cite{Hanahan2000Hakkmarks,
Hanahan2011Hakkmarks, Fouad2011Revisiting}. Cancer arises through a multistep, mutagenic process whereby cancer cells acquire a common set of
properties including unlimited proliferation potential, self-sufficiency in growth signals, and resistance to antiproliferative and apoptotic
cues. Tumors evolve to garner support from surrounding stromal cells, attract new blood vessels to bring nutrients and oxygen, evade immune
detection, and ultimately metastasize to distant organs.

Some modern papers define the following seven hallmarks of cancer: selective growth and proliferative advantage, altered stress response
favoring overall survival, vascularization, invasion and metastasis, metabolic rewiring, an abetting microenvironment, and immune modulation
\cite{Hanahan2000Hakkmarks, Hanahan2011Hakkmarks, Fouad2011Revisiting}. An example is RAS protein. It is chronically active in $30\%$ of cancers
and in over $90\%$ of pancreative carcinomas, often via missense
mutations in its gene or inactivating mutations in one of its negative regulators (e.g. NF1). Mutations in RAS result in a plethora of
effects beyond enhanced growth and proliferation that include suppression of apoptosis, rewiring of metabolism, promoting angiogenesis, and 
immune evasion. A single signaling cascade could be implicated in multiple hallmarks of cancer. Exactly how cancer cells escape dormancy
(``awaken'') and  proceed to form overt metastasis (colonization) is poorly understood!

We believe that parameters $q$, $K$ and $X_{\infty}$ are related to evading apoptosis, insensitivity to anti-growth signals, sustaining
proliferative signaling and chronic proliferation. Parameter $X_{\infty}$ is connected with evading growth suppressors, enabling replicative
immortality, limitless replicative potential, genome instability, and rewiring metabolism. Parameters $d$, $e$, $f$, $V$, $b$ and 
$X_{\infty}$ are associated with the immune system, rewiring metabolism, angiogenesis, microenvironment, inflammatory cells, tissue invasion, and 
metastasis. Parameter $q$ is linked to oncogene addiction. In particular, $q$ can be changed targeting cancer stem cells, oncogenes, and
tumor-suppressor genes. Figure \ref{Hallmarks}b show some connections between the parameters and the Hallmarks of cancer. Tissue invasion and
metastasis can be affected by parameter $X_{\infty}$. By targeting individual hallmarks features or enabling characteristics it may be possible
to achieve therapeutic benefit. We should attack as many hallmarks as possible.

Experiments have corroborated the existence of immune surveillance. Despite immune
surveillance, tumors continue to develop in bodies with intact immune systems. Cancer immunoediting is the process by which the immune system eliminates and shapes malignant
disease and encompasses three phases: \emph{elimination}, \emph{equilibrium}, and \emph{escape}. Inactivation of the second p53 allele leads to increased
cell proliferation,  decreased apoptosis, and tumor development \cite{bressy2017combining}. BRCA1 and BRCA2 are tumor suppressor genes associated 
with breast and ovarian cancers, along with several other cancers.

HER2/neu is overexpressed in $25\%$ of breast cancers. These cancers tend to be more aggressive
clinically. RAS is the oncogene most commonly activated in human tumors. The human genome encodes three RAS genes: H-ras, K-ras, and N-ras.
A large fraction of tumors contain mutations in one of these three genes. For example, $70-90\%$ of pancreatic carcinoma contain a mutation
in the K-ras gene. The MYC oncogene is often amplified or overexpressed in cancers.

Many oncogenes are responsible for the self-sufficiency in Growth Signals of the cells. For example, RAS is found mutated in about $25\%$ of
human tumors, thus leading to ligand-independent activation of the ras-raf-mapk signaling pathway. Insensitivity to antigrowth signals is
produced by the loss of tumor suppressor genes.

Cancer cells evade antiproliferative signals. At the molecular level, many anti-proliferative signals are funneled through the Rb
protein (and its two relatives, P107 and P130). Antigrowth signals are impaired e. g. by the loss of CDK inhibitors (P16, P21, p53). 
Resistance to apoptosis can be acquired by cancer cells through a variety of strategies. For example, cancer cells can evade apoptosis by the
loss of p53.

Self-sufficiency in growth signals, insensitivity to antigrowth signals does not ensure expansive tumor growth. Most cells possess a mechanism
that limits their proliferation using the number of telomere repeats at the end of chromosomes. Some tumors cells avoid this via a mutation
that upregulates expression of the telomerase enzyme, thus acquiring limitless replicative potential by maintaining their telomeres.


\section{Therapy strategies}
\label{Sec:Strategies}

This research work has proved seminal in rational drug design. The main strategy is to increase $q$ such that $q>1$. In this case, the cancer is
curable because the fixed point $P_I$ can, in principle, be stable.
Of course, this does not guarantee that the point $P_I$ is stable. For instance, for $q=2$, we need in addition $Vb>af$. Thus,
immunotherapy plays an important role.

Equation \ref{Eq03} is related to the microenvironment. Anti-angiogenesis can limit the expansion of the tumor. Therapies that reduce
$X_{\infty}$ can eliminate the limitless proliferation and many other cancer-related properties. A very small $X_{\infty}$ can lead to 
complete elimination of cancer (see Fig.~\ref{Fig01}e).
Some therapy strategies are:

\begin{enumerate}
 \item Oncogene-targeted and tumor suppressor targeted therapies (sequential, in the sense that it should be applied before the combinations
 described later).
 
 \item Immunotherapy, anti-angiogenesis, oncogene-targeted chemotherapy (concurrent).
 
 \item Immunotherapy, oncolytic virotherapy, plus transgene p53, plus chemo-radiation, logarithmic cell-kill function (concurrent).
\end{enumerate}

Below we discuss several strategies including connections between Hallmarks and combinations \cite{Wang2016Combination}. Combination therapy
targeting both cancer stem-like cells and bulk tumor cells should improve the efficacy of breast cancer treatment. The cancer stem-like cells
hypothesis suggests that tumor development and metastasis are driven by a minority population of cells, which are responsible for tumor
initiation, growth, and recurrence. The inability to efficiently eliminate cancer stem-like cells during chemotherapy, together with these
cells being highly tumorigenic and invasive, may result in treatment failure due to cancer relapse and metastases.

The ideal panacea for cancer would kill all malignant cells, including cancer stem-like cells and bulk tumor cells. Both chemotherapy and cancer
stem-like cells are insufficient to cure cancer. We need combination therapy with \emph{cancer stem-like cells-targeted} agents and 
chemotherapeutics.

\emph{Salinomycin} plays an important role against the cancer stem-like cells. Combination therapies of Salinomycin with conventional
chemotherapy (paclitaxel or lipodox) showed a potential to improve cell killing.
Cancer stem-like cells are highly resistant to standard chemo- and radiotherapies and they persist following treatment. Conventional
therapies may achieve clinical tumor reduction at the beginning of treatment but are insufficient to cure cancers \cite{McDermott2010Targeting}.
Thriving cancer stem-like cells under the influence of the stem cell microenvironment lead to tumor recurrence and disease relapse and 
even promote the formation of distant metastases, ultimately leading to treatment failure following chemo- and radiotherapy \cite{Hart1998,
Calderon1991, Stepanova1979, gonzalez2003}.

The combinatorial therapy of paclitaxel, lipodox, and salinomycin produced a synergetic effect and led to a significantly higher cytotoxicity
than the use of either lipodox or salinomycin alone at all concentrations and time points tested. A synergy between paclitaxel and
salinomycin for growth inhibition were seen with a long term drug incubation for 72 hours in all 6 combination treatment groups, indicating
that combination therapy enhanced antitumor activity if \emph{prolonged exposure of drugs was used}.

Eradication of all malignant cells within a patient's cancer including cancer stem-like cells and their progeny is essential to prevent
cancer relapse and metastasis. Standard chemo- and radiotherapy may have clinical benefits on tumor regression in advanced stages of cancer
as a result of their killing the bulk tumor population, but disease relapse is highly likely to occur due mainly to their minimal effect on
the cancer stem-like cells population. A cancer stem-like cells-targeted therapy may have substantial clinical benefit.

We would like to stress that cancer stem cells, oncogene addiction, and tumor suppressor gene activity are related to parameter $q$.
Accumulated evidence supports the conclusion that cancer stem-like-cells-targeting agents are most effective in eradicating cancer when these
agents are in combination with conventional cytostatic drugs and/or novel cancer stem-like cell-targeted drugs \cite{Cattani2008, Ledzewicz2013,
Wilkie2013}. Targeted strategies being developed include direct inhibition of the self-renewal of stem-like cells, indirect modulation of the
microenvironment and direct induction of death of cancer stem cells by chemical agents that trigger differentiation of cancer stem-like
cells, immunotherapy, and oncolytic viruses \cite{Phuc2012}.

Oncogene addiction refers to the curious observation that a tumor cell, despite its plethora of genetic alterations, can seemingly exhibit
dependence on a single oncogenic pathway or protein  of its sustained proliferation and/or survival. A profound implication of this
hypothesis is that switching off this crucial pathway upon which cancer cells have become dependent should have devastating effects on the 
cancer cell while sparing normal cells that are not similarly addicted. This is the discriminating activity required for effective cancer
therapeutic. 

Activated kinases are the Achilles heel of many cancers. Just as acute inactivation of addicting oncoproteins frequently leads to cancer cell
death, recent evidence points to similar outcomes engendered by the reintroduction of wild-type versions of tumor suppressor genes that are
frequently inactivated in cancer cells. Tumor suppressor hypersensitivity may represent another dimension of oncogene addiction since it is
likely that in the establishment of oncogene addicted state, a prerequisite may involve the removal of support systems such as tumor
suppressors that buttress normal cell survival.

MYC-induced apoptosis plays a physiological role in antigen-induced negative selection of developing T cells. An important and
therapeutically relevant feature of oncogene addiction relates to the tumor cell-specific induction of apoptosis. This phenomenon is related
to parameter $q$. Acute oncoprotein inactivation in cancer cells often results in apoptosis. Factors that influence apoptosis are likely
to influence the response to oncoprotein inactivation in oncogene inactivation in oncogene addicted tumors. Dysregulated cell proliferation
and suppression of apoptosis are hallmarks of cancer. Cancer cells deploy a diverse array of mechanisms to avoid apoptosis, such as
inhibition of cell death mediating proteins, and/or overexpression of cell death inhibitory proteins.

The oncogenic shock model may also have implications regarding \emph{optimal strategy for treating patients} with targeted inhibitors. The 
oncogenic shock model may also have implications for the use of drug combinations in cancer therapy. The correct design of cancer
chemotherapy drugs schedules is very important here. Oncogene addiction may lead to new therapeutic strategies. The identification of
addiction settings involving codependence on two different oncogene products may lead to a rational combination treatment strategy involving
the simultaneous disruption of both genes.

The phenomenon of tumor suppressor hypersensitivity could also lead to a therapeutic opportunity in which a tumor suppressor is reintroduced
into cancer cells. Cancer cells growth and survival can often be impaired by the inactivation of a single oncogene. For this, we have to
identify the Achilles heel. \emph{Combinational therapy} is required to prevent the escape of cancers from a given state of oncogene
addiction. It is unlikely that the use of a single molecular targeted agent will achieve long-lasting remissions on cures in human cancers, 
especially for late-stage disease. Combinational therapy will, therefore, be required. This combination should be \emph{rationally designed}.

Clinical studies \cite{Weinstein2006Mechanism} indicate that the efficacy of certain molecularly targeted agents can be enhanced by
combining them with cytotoxic agents, i. e. agents that often act by inhibiting DNA or chromosomal replication. In order to understand how 
our strategies are confirmed by experimental and clinical studies, see the References \cite{Wang2016Combination, Weinstein2006Mechanism,
Pagliarini2015Oncogene, Hijikata2018Phase, Mahadevan2014Misc, Torti2011Oncogene, Fischer2017Census, Guest2016Functional,
Deraedt2011Exploiting, Lasalvia-Prisco2012Addition, bozic2013evolutionary,  sneddon2007location, yang2008nf1, hurwitz2004bevacizumab,
lim2017mechanisms, brahmer2012safety, rak1995mutant, cairns2006overcoming, fan2003combinatorial, feldmann2007blockade, radpour2017tracing,
grossenbacher2016natural, Miller2005, ren2014anti, weinstein2008oncogene, sharma2007oncogene, hu2014combining, gerber2005pharmacology,
simpson2016cancer, chaurasiya2018oncolytic, gatzka2018targeted, lin2018advances, polivka2017advances, allen2017combined, el2016combined,
lawson2018oncogenic, bressy2017combining, idema2007addelta24}.

Trastuzumab that targets HER2 can improve response and survival rates if given in combination with paclitaxel to patients with metastatic
breast cancer. The combination of bevacizumab or cetuximab with cytotoxic chemotherapy agents can also improve response rates in patients
with metastatic breast and colon cancer. When bevacizumab was added to a combination chemotherapy regime, it improved overall survival rates
in patients with metastatic colon cancer.

We have applied experimental design techniques \emph{a posteriori} as an inverse problem using data from experiments and clinical studies.
The conclusion of this analysis is that the combination of therapies that we propose corroborate the mathematical model results and provide
additional evidence for the therapy schedule that we have designed.

\section{Discussion}
\label{Sec:Discussion}

Our objective with these models is to achieve Hanahan's and Weinberg's dream: to develop cancer research into a logical science
\cite{Hanahan2000Hakkmarks}. ``Cancer 
prognosis and treatment will become a rational science. It will be possible to understand with precision how and why treatment regimens
and specific antitumor drugs succeed or fail''. In this spirit, we can develop their ideas in such a way that anticancer drugs are targeted
to each of the hallmark capabilities of cancer. Some, used in appropriate combinations, will be able to prevent incipient cancers from 
developing, while others will cure pre-existing cancers, elusive goals at present. Dynamical systems can help cancer biology to become a
science with the mathematical structure and logical consistency similar to that of physics.

The outcome of dynamical system (\ref{Eq02}-\ref{Eq03}) depends on the set of parameters and the initial conditions. For example, in the
phase space represented in Fig.~\ref{Fig01}a (for a fixed set of parameters), the initial conditions decide everything. For initial conditions
given by points $\alpha$, $\beta$, and $\delta$, the phase trajectories are attracted to the fixed point $P_I$, where $X=0$. This is a good outcome.
Notice that for the initial point $\delta$, the size of the tumor is large. However, the dynamics lead to a cure. All the mentioned points
($\alpha$, $\beta$, and $\delta$) are to the left of the separatrix of saddle point $P_{II}$. On the other hand, the fate of points $\gamma$ and 
$\varepsilon$ is fatal. These points are to the right of the separatrix of the fixed point $P_{II}$.

In the phase space shown in Fig.~\ref{Fig01}b, the fixed point $P_I$ is unstable. There is an additional stable point $P_{III}$ with $X\neq0$.
This point represents a dormant state. The initial conditions defined by the points $\zeta$ and $\eta$ will lead to phase trajectories that are
attracted to the point $P_{III}$. However, other initial data (like point $\theta$) will lead to an unlimited increase of the tumor size.

In the phase space shown in Fig.~\ref{Fig01}c, the only fixed point $P_I$ is unstable and all the phase trajectories lead to an unlimited
increase of the tumor. Fig.~\ref{Fig01}d shows a phase space where $X_{\infty}\neq\infty$. Note that there is a stable fixed point $P_{IV}$, that
is now the attractor that symbolizes the limit maximum size of the tumor. The initial data that stand for the points $\varphi$ and $\mu$,
will go to the point $P_I$ ($X=0$), while the point $\psi$ will move to the fixed point $P_{IV}$. The latter is not a good prognosis.

When $X_{\infty}$ is small, the outcome can be very favorable. This is shown in Fig.~\ref{Fig01}e. $X_{\infty}$ is related to many cancer
hallmarks and genes. If $X_{\infty}$ is finite, there is no limitless proliferation. The invasion is also bounded. A small $X_{\infty}$ is
an evidence that the microenvironment is not friendly. Fig.~\ref{Fig01}e shows the dynamics when $V/f>a/b$ and
\begin{equation}
X_{\infty}<\frac{11}{6}\frac{1}{e} +\frac{2}{3}\frac{f}{d}.
\end{equation}
Note that all the phase trajectories tend to the fixed point $P_I$ ($X=0$).

Therapy can change the outcome. Let us return to Fig.~\ref{Fig01}a. The initial point $\gamma$ would lead to a fatal outcome. However, even a slight
change in the constant therapy $-C_oX$ in the system (\ref{Eq02}-\ref{Eq03}), can change parameter $a=2K$. This can conduct to a reposition
of the separatrix of fixed point $P_{II}$ such that the $\gamma$ point will ride a trajectory moving to the point $P_I$ ($X=0$).

We have already discussed the unfortunate situation when $q\leq1$, which provokes the fixed point $P_I$ to be unstable. There is a therapy that
can change $q$ (see previous sections). In that case, we can be in a situation like the one shown in Fig.~\ref{Fig01}a. Both cell-killing
therapies and immunotherapy can get the dynamics on a trajectory moving to a (now stable) fixed point $P_I$ ($X=0$). The articles
\cite{Wang2016Combination,
Weinstein2006Mechanism, Pagliarini2015Oncogene, Hijikata2018Phase, Mahadevan2014Misc, Torti2011Oncogene, Fischer2017Census, Guest2016Functional,
Deraedt2011Exploiting, Lasalvia-Prisco2012Addition, bozic2013evolutionary, sneddon2007location, yang2008nf1, hurwitz2004bevacizumab,
lim2017mechanisms, brahmer2012safety, rak1995mutant, cairns2006overcoming, fan2003combinatorial, feldmann2007blockade, radpour2017tracing,
grossenbacher2016natural, Miller2005, ren2014anti, weinstein2008oncogene, sharma2007oncogene, hu2014combining, gerber2005pharmacology,
simpson2016cancer, chaurasiya2018oncolytic, gatzka2018targeted, lin2018advances, polivka2017advances, allen2017combined, el2016combined,
lawson2018oncogenic, bressy2017combining, idema2007addelta24} corroborate these statements. 

\begin{figure}
\begin{center}
\scalebox{0.35}{\includegraphics{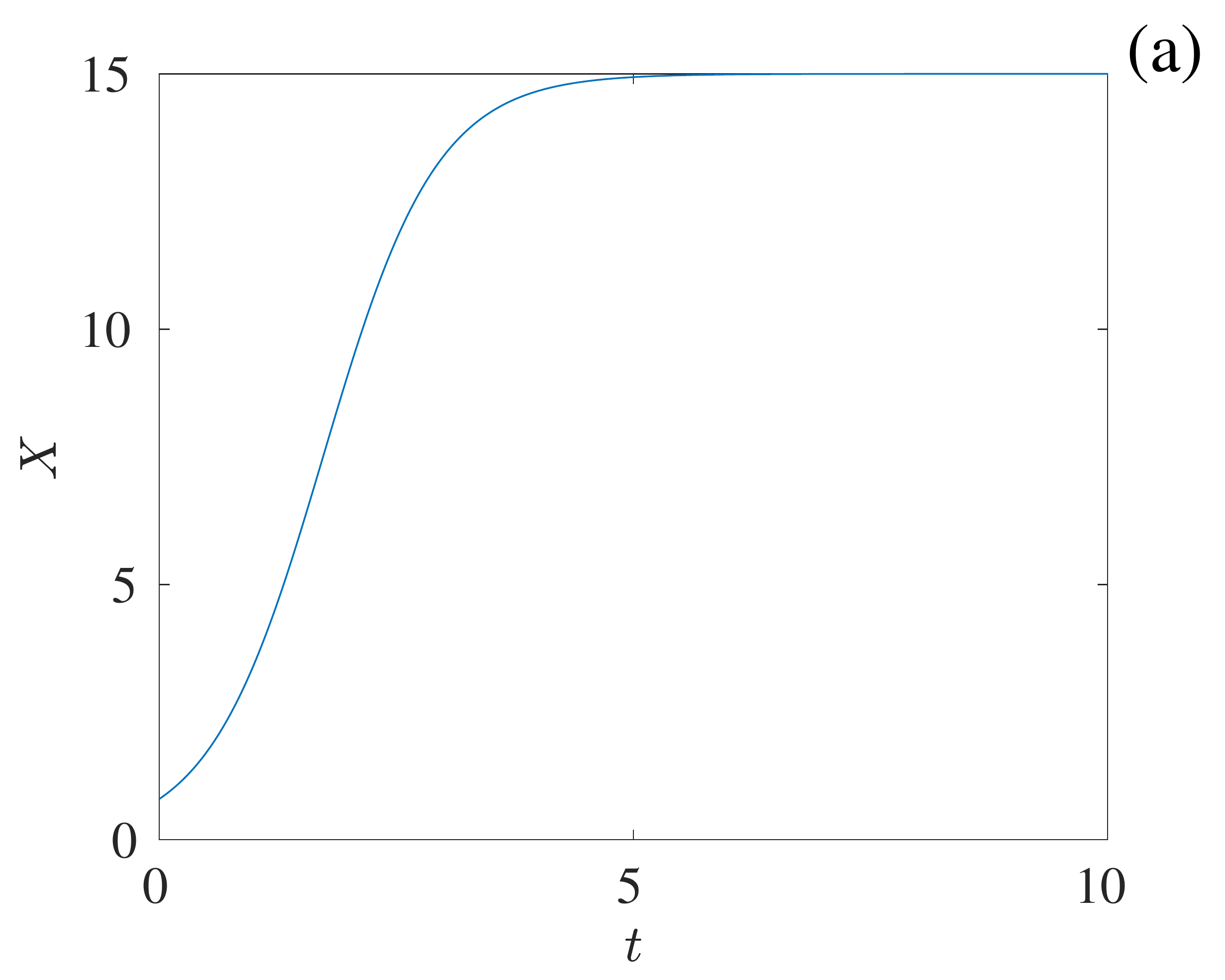}}\scalebox{0.35}{\includegraphics{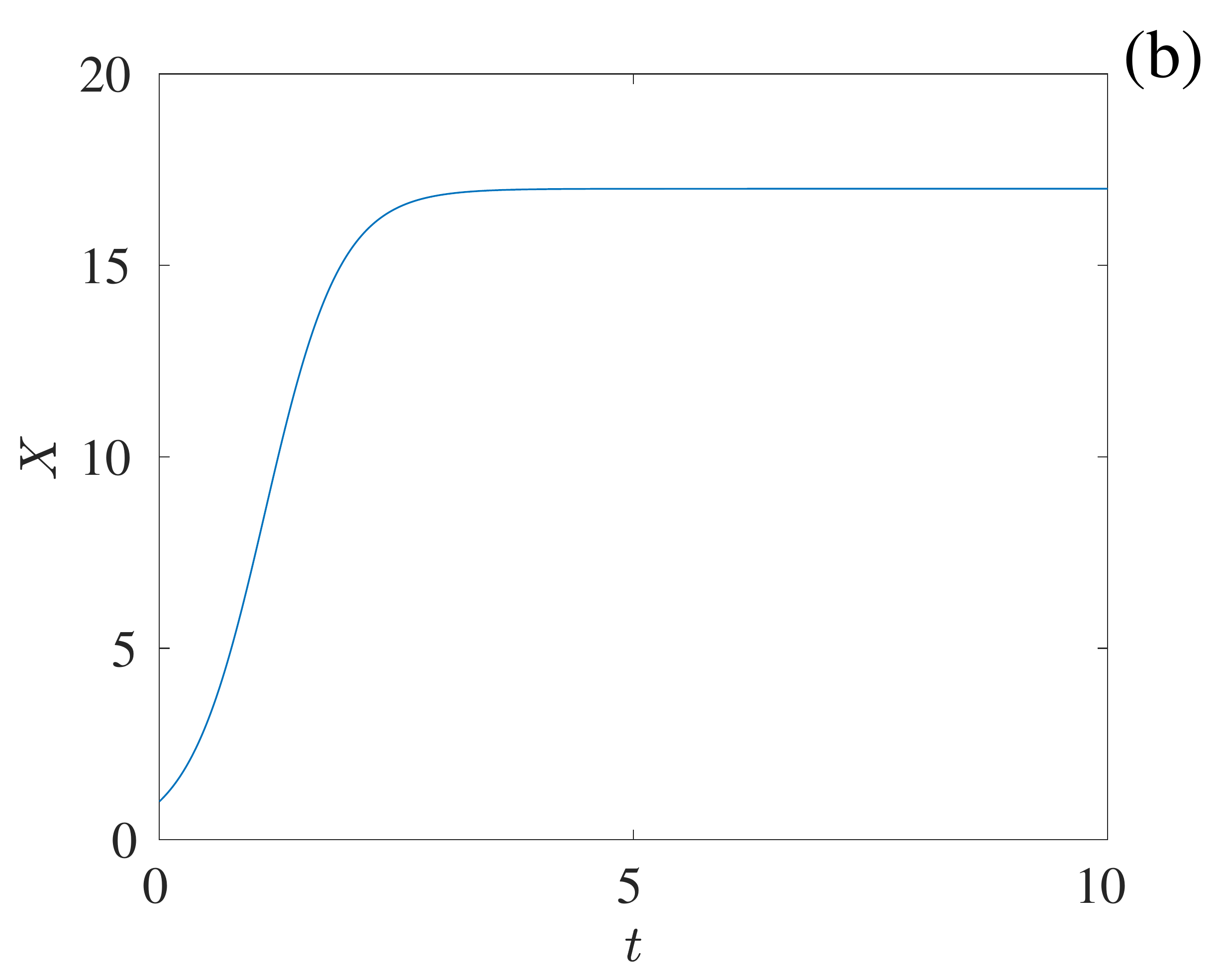}}
\scalebox{0.35}{\includegraphics{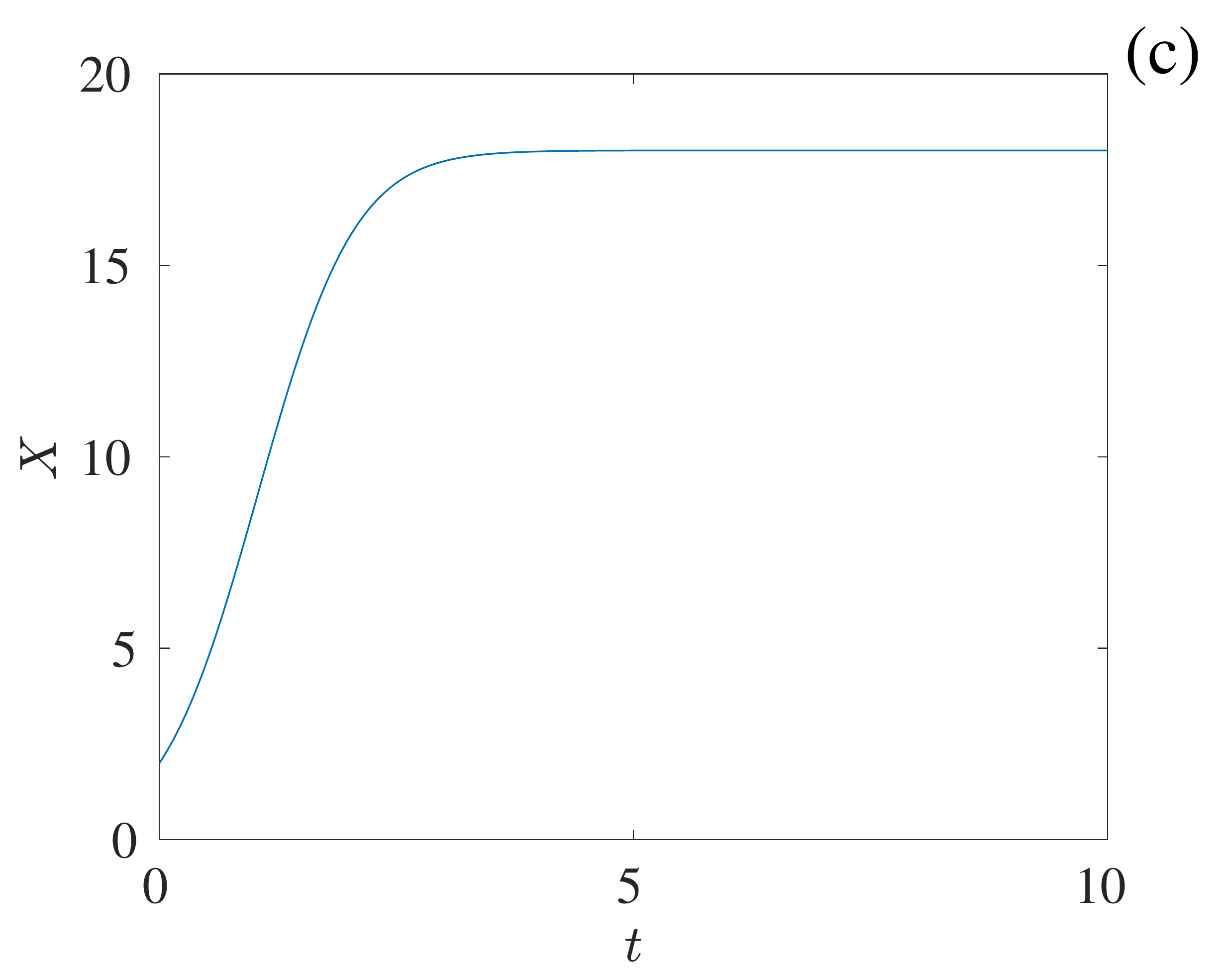}}
\end{center}
 \caption{Behavior of $X(t)$ in Eq.~(\ref{Eq01}). Each case shows different outcomes for different parameters. \textbf{(a)} $K=1$, $X_{\infty}=15$, and $q=2$.
 \textbf{(b)} $K=2$, $X_{\infty}=17$, and $q=3$.
 \textbf{(c)} $K=2$, $X_{\infty}=18$, and $q=3/2$. \label{Fig03}}
\end{figure}

 \begin{figure}
 \begin{center}
 \scalebox{0.35}{\includegraphics{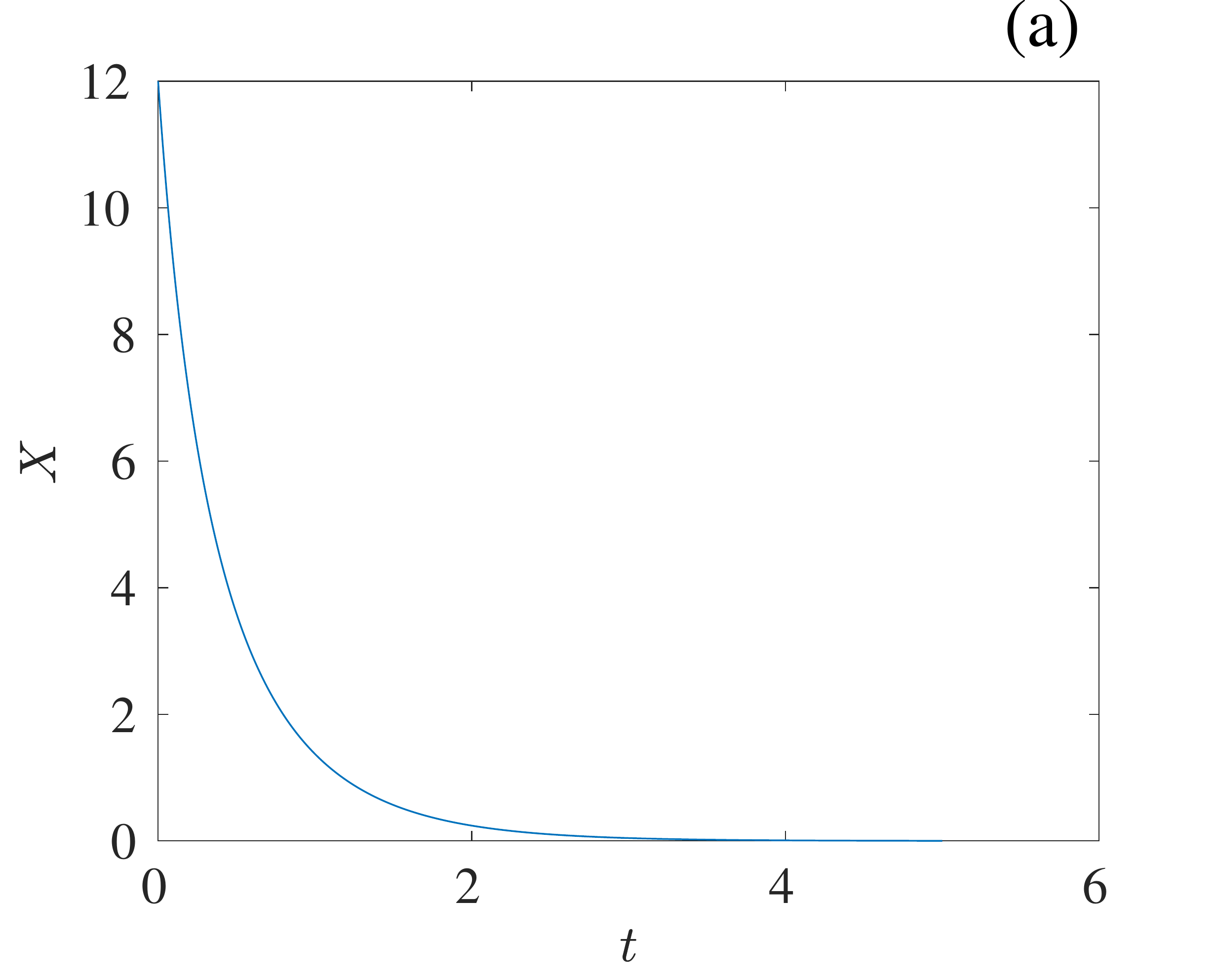}}\scalebox{0.35}{\includegraphics{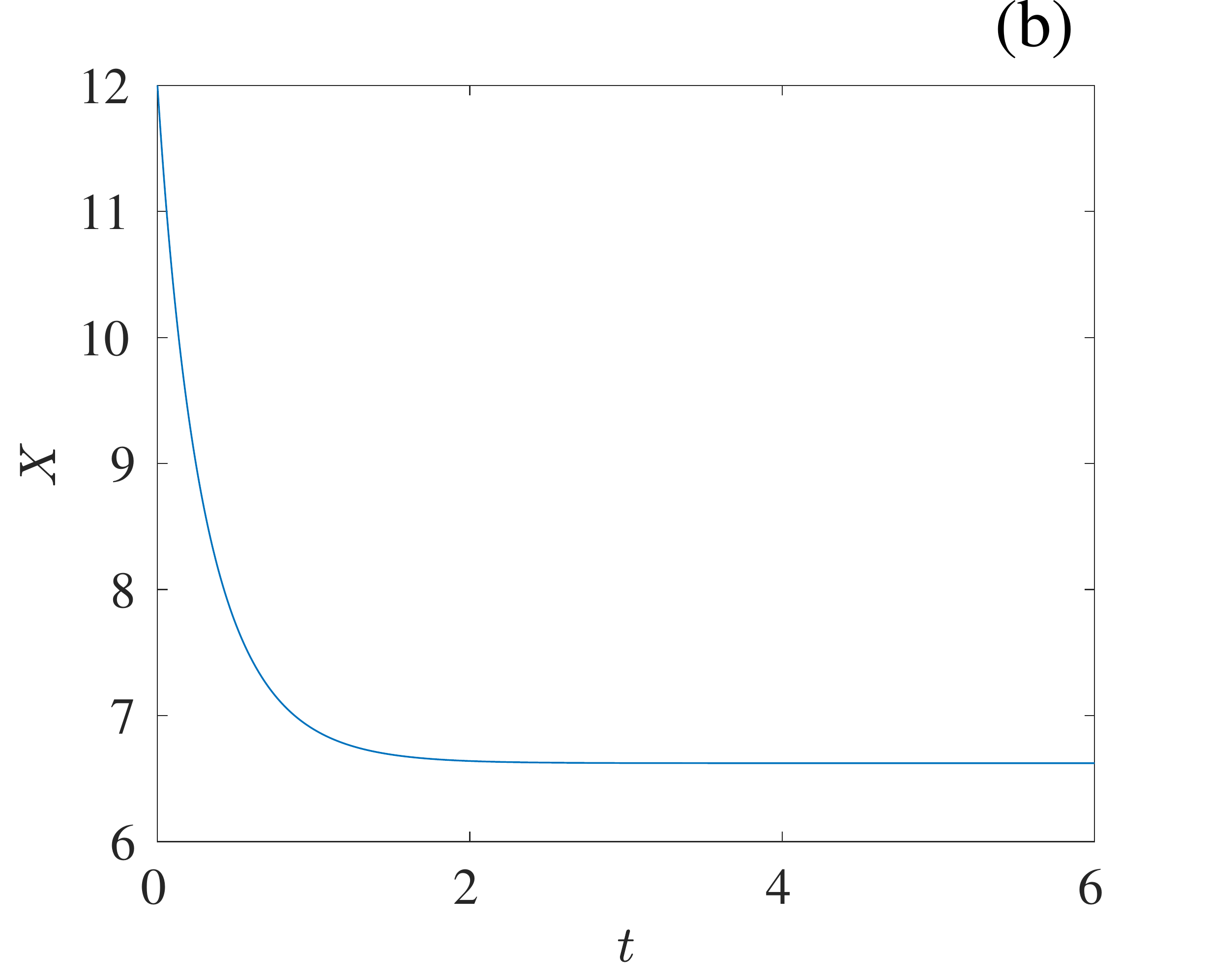}}
 \scalebox{0.35}{\includegraphics{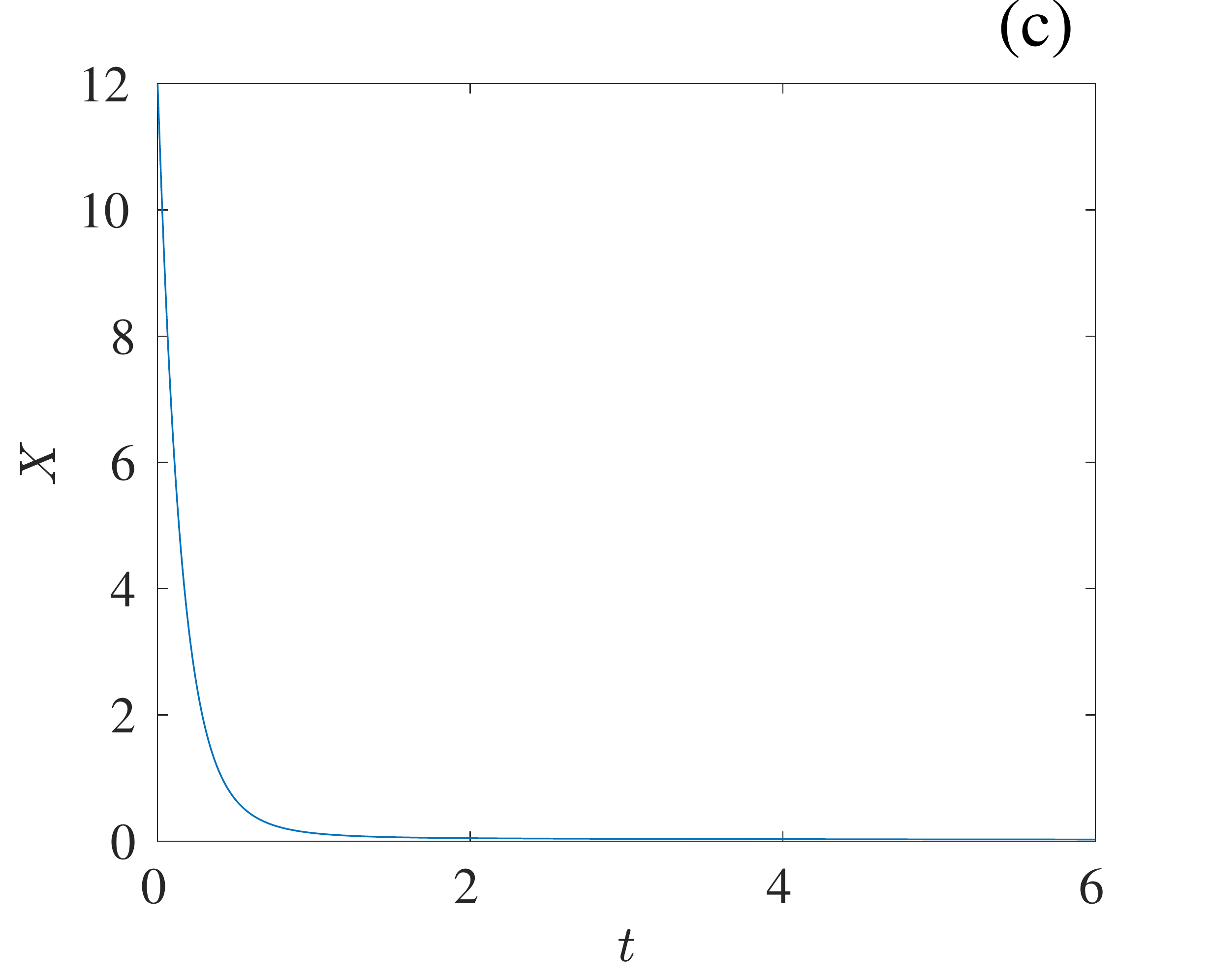}}\scalebox{0.35}{\includegraphics{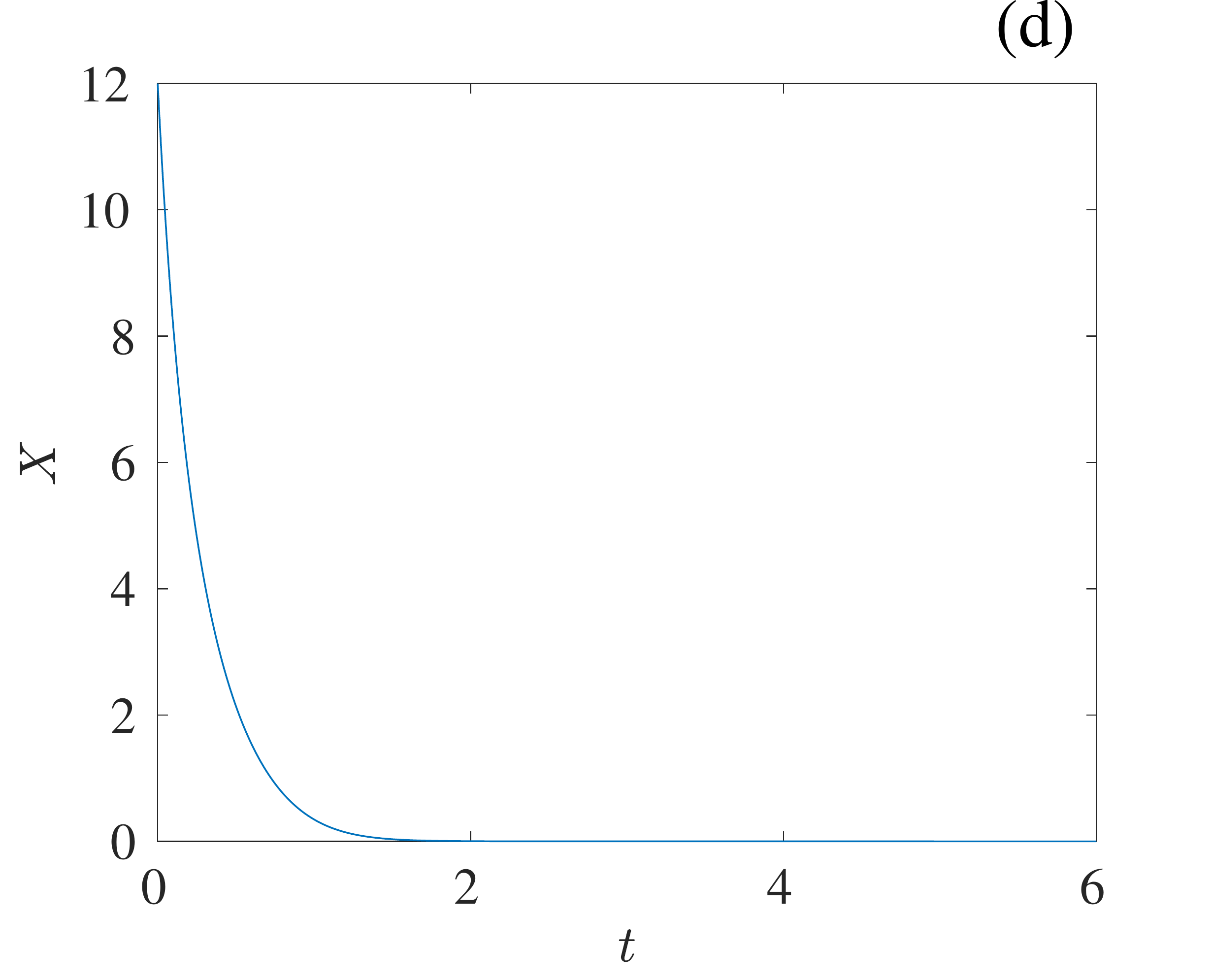}}
 \end{center}
  \caption{The response of cancer in the presence of different parameters and therapies. \textbf{(a)} $K=1$, $X_{\infty}=15$, $q=2$, and $C(t)=4$.
  Note that in this case $q>1$. So we can find a constant dose $C(t)=C_o$ able to change parameter $a$ in such a way that the tumor will be
  reduced to zero.
  \textbf{(b)} $K=2$, $X_{\infty}=17$, $q=0.6$, and $C(t)=4$. Note that $q<1$. A conventional therapy will never eradicate the cancer.
  \textbf{(c)} $K=1$, $X_{\infty}=18$, $q=0.8$, and $C(t)=6(1+t^{0.2})$. Although $q<1$, a late-intensification schedule can eradicate the
  cancer. \textbf{(d)} $K=1$, $X_{\infty}=18$, $q=0.8$, and $C(t)=6(1+t^2)$. Another example of late-intensification schedule capable of
  eliminating the cancer.
  \label{Fig04}}
 \end{figure}

\subsection{Treatment schedules}

Even in our previous papers \cite{deVladar2004, gonzalez2003, gonzalez2006}, we had pointed out the importance of combination therapies and 
time-dependent treatments. Figure \ref{Fig03} shows the behavior of the untreated tumors according to Eq.~(\ref{Eq01}). This is exactly what
happens in real life. When $q>1$, and the host response is strong enough, we can reduce the tumor size to zero even with constant therapy. This
is shown in Fig.~\ref{Fig04}a. If $q=0$ or $q<1$, the cancer cannot be cured using conventional constant-dose therapy. We can observe this in
Fig.~\ref{Fig04}b.

We have discussed late-intensification schedules in other articles \cite{deVladar2004, gonzalez2003, gonzalez2006}. Many regular physicians are
considering the logarithmic treatment as the optimal therapy in their practice with real patients. Indeed, our logarithmic therapy has been studied
by other scientists and physicians with great success (see e.g. \cite{sahoo2011stochastic, Lo2009}). In Ref.~\cite{Lo2009}, the
author has found that ``during cancer treatments the dose intensity should not be decreased at any time because this will allow the tumor to
relapse, and the logarithmic intensification therapy could be the optimal therapeutic strategy''. A similar statement can be found in
Ref.~\cite{sahoo2011stochastic}. Compare this with the results shown in this article and in Ref.~\cite{deVladar2004, gonzalez2003, gonzalez2006}. 

We are including late-intensification in the new combination therapies that we are designing. We are finding that these new strategies can be even
more powerful than the previously employed treatments. Physicians can face drug resistance during chemo. Here we present some examples of other
ways to apply late-intensification therapies.

\begin{itemize}
 \item Increasing the number of different drugs in the combination following a logarithmic function.
 
 \item Changing the types of drugs during late-intensification.
 
 \item Using new, more precise forms of applying radiation: electrons, protons, etc. This allows for the late-intensification to be applied
 without increasing the danger to normal cells.
 
 \item Increasing the number of viruses in virotherapy.
\end{itemize}

\section{Conclusions}
\label{Sec:Conclusions}

Most cancer tumors are generally untreatable using conventional therapies. Combination of anti-angiogenesis with chemotherapy leads to more
effective cancer treatment. Synergistic antitumor activity results when metronomic cyclophosphamide is combined with a tumor-targeted 
immunotherapy. The repair or inactivation of mutant genes may be effective in treatment of cancer. Drugs that target oncogenes have been
shown to be effective in the treatment of some cancers. Although many of the new drugs show a significant clinical response, eventually most
tumors do reoccur. Combination therapies are needed. We need therapies with the potential to target both tumor cells and the tumor
microenvironment. Nevertheless, tumor cells posses the ability to circumvent most therapeutic agents when given as monotherapy.

We have found that the success of the new therapeutic agents is seen when used in combination with other kinds of therapies, including
conventional treatments. Among many genetic lesions, mutational inactivation of the p53 tumor suppressor gene is one of the most frequent
events found in human cancers. Tumor suppressor gene p53 plays a critical role in tumor progression, mainly by inducing growth arrest,
apoptosis, and senescence, as well as by blocking angiogenesis. In addition, p53 generally confers the cancer cell sensitivity to 
chemoradiation. The successful development of p53 modulator drugs (to be used in combination with immunotherapy and chemo-radiation) can
change current anticancer therapies.

Using our mathematical models and data from experiments and clinical studies, we have designed the following sequence of several sets of
combinational therapies.

\begin{enumerate}
  \item Oncogene-targeted therapy, tumor suppressor targeted therapy, and killing of cancer stem-like cells.
 
 \item Immunotherapy, anti-angiogenesis, oncogene-targeted therapy, and chemotherapy. The goal of this round is to make the fixed
 point $P_I$ stable. These treatments are supposed to be applied concurrently. The conditions that arise following these treatments should make
 possible the eradication of cancer.

 \item Immunotherapy, Oncolytic Virotherapy plus transgene p53, chemo-radiation, and logarithmic cell-kill function (concurrent application).
 This round should move the system to the left part of separatrix of the saddle fixed point $P_{II}$.

\end{enumerate}

\section*{Acknowledgements}

M.A.G-N. thanks for the financial support of Proyecto Interno Regular PUCV 039.306/2018. J.F.M. acknowledges the financial support of CONICYT
doctorado nacional No. 21150292.

\bibliographystyle{unsrt}

\end{document}